\documentclass{aa}
\usepackage{graphicx}
\usepackage{lscape}
\usepackage{subfigure}
\usepackage{natbib}
\usepackage[varg]{txfonts}
\usepackage{longtable}
\usepackage{booktabs}
\usepackage[normalem]{ulem} %added temporarily to cross-out sentences
\usepackage{epstopdf}
\usepackage{xcolor}
\usepackage{soul}
\usepackage{makecell}
\usepackage{multicol,tabularx,capt-of}

\usepackage{scalerel}
\usepackage{tikz}
\usetikzlibrary{svg.path}

\defcitealias{Sanchez-Saez2023}{SS23}

\definecolor{orcidlogocol}{HTML}{A6CE39}
\tikzset{
  orcidlogo/.pic={
    \fill[orcidlogocol] svg{M256,128c0,70.7-57.3,128-128,128C57.3,256,0,198.7,0,128C0,57.3,57.3,0,128,0C198.7,0,256,57.3,256,128z};
    \fill[white] svg{M86.3,186.2H70.9V79.1h15.4v48.4V186.2z}
                 svg{M108.9,79.1h41.6c39.6,0,57,28.3,57,53.6c0,27.5-21.5,53.6-56.8,53.6h-41.8V79.1z M124.3,172.4h24.5c34.9,0,42.9-26.5,42.9-39.7c0-21.5-13.7-39.7-43.7-39.7h-23.7V172.4z}
                 svg{M88.7,56.8c0,5.5-4.5,10.1-10.1,10.1c-5.6,0-10.1-4.6-10.1-10.1c0-5.6,4.5-10.1,10.1-10.1C84.2,46.7,88.7,51.3,88.7,56.8z};
  }
}

\newcommand\orcidicon[1]{\href{https://orcid.org/#1}{\mbox{\scalerel*{
\begin{tikzpicture}[yscale=-1,transform shape]
\pic{orcidlogo};
\end{tikzpicture}
}{|}}}}

\usepackage{hyperref} %<--- Load after everything else
\hypersetup{
    colorlinks=true,
    citecolor=blue,
    linkcolor=blue,
    urlcolor=blue,
    }
\makeatletter
\renewcommand*\aa@pageof{, page \thepage{} of \pageref*{LastPage}}
\makeatother

\bibpunct{(}{)}{;}{a}{}{,} % to follow the A&A style
\usepackage{amstext} 
\definecolor{cadmiumred}{rgb}{0.89, 0.0, 0.13}

\definecolor{ste}{rgb}{0., 0.26, 0.15}

\begin{document} 
   \title{Unlocking AGN Variability with Custom ZTF Photometry for High-Fidelity Light Curves and Robust Selection}

   \author{
   P. Ar\'evalo\inst{1,2,3\orcidicon{0000-0001-8503-9809}}, P. S\'anchez-S\'aez\inst{4\orcidicon{0000-0003-0820-4692}},
   and B. Sotomayor\inst{1}, P. Lira\inst{5,2}, F. E. Bauer\inst{6\orcidicon{0000-0002-8686-8737}}, S. Ríos\inst{5,2}}

   \titlerunning{Custom ZTF Photometry for High-Fidelity Light Curves}
   \authorrunning{Ar\'evalo et al.}

\institute{
Instituto de F\'isica y Astronom\'ia, Facultad de Ciencias, Universidad de Valpara\'iso, Gran Breta\~na 1111, Valpara\'iso, Chile
\\e-mail: patricia.arevalo@uv.cl
\and
Millennium Nucleus on Transversal Research and Technology to Explore Supermassive Black Holes (TITANS)
\and Millennium Institute of Astrophysics (MAS), Nuncio Monseñor Sótero Sanz 100, Providencia, Santiago, Chile
\and
European Southern Observatory, Karl-Schwarzschild-Strasse 2, 85748 Garching bei München, Germany
\\e-mail: pasanchezsaez@gmail.com, paula.sanchezsaez@eso.org
 \and Departamento de Astronom{\'{\i}}a, Universidad de Chile, Camino el Observatorio 1515, Santiago, Chile
 \and Instituto de Alta Investigaci{\'{o}}n, Universidad de Tarapac{\'{a}}, Casilla 7D, Arica, Chile}

   \date{}
  \abstract
   {}
   {We explore the potential of optical variability selection methods to identify a broad range of AGN, including those challenging to detect with conventional techniques. Using the unprecedented combination of depth, sky coverage, and cadence of the Zwicky Transient Facility (ZTF) survey, we specifically target low-luminosity, low-mass, and starlight-dominated AGN, known for their redder colours, weaker variability signals, and difficult nuclear photometry due to their resolved host galaxies.}
   {We perform aperture photometry on ZTF reference-subtracted $g$-band images for $\approx$39.8 million sources across $>$8,000 deg$^2$, assemble light curves and calculate features for all detected sources, and classify objects employing a Random Forest algorithm into 14 distinct classes, including 341,938 candidate AGN across four classes (low-z, mid-z, high-z, and blazars). We compare variability metrics derived from our photometry to those obtained from publicly available ZTF Data Release light curves, obtained through psf-photometry on the science, i.e. not reference-subtracted images (DR11-psf), to assess the impact of our analysis. 
   Finally, we compare our AGN candidate sample with those identified through colour selection and X-ray detection techniques.
}
   {
   We find that the fraction of low-z quiescent galaxies exhibiting significant variability drops dramatically (from 98\% of the sample to 7\% of the sample, when using the standard variability metric Pvar) when replacing the DR11-psf light curves with our difference image, aperture photometry (DI-Ap) version. The overall number of variable low-z AGN remains high (99\% when using DR11-psf lightcurves, 83\% when using DI-Ap), however, implying that our photometry can detect the fainter variability in host dominated AGN. The classifier effectively distinguishes between AGN and other sources, demonstrating high recovery rates even for AGN in resolved nearby galaxies. 
   AGN candidates in eROSITA's eFEDS field, detected in X-rays and bright enough for ZTF optical observations, were classified as AGN (79\%) and non-variable galaxies (20\%). These groups show a 2 dex difference in X-ray luminosity but not in X-ray flux. A significant fraction of X-ray AGN are optically too faint for ZTF, and conversely, a quarter of ZTF AGN in the eFEDS area lack X-ray detections, highlighting a wide range of X-ray-to-optical flux ratios in the AGN population. }
   {}

   \keywords{galaxies: active -- surveys -- methods: statistical -- methods: data analysis }
   \maketitle

\section{Introduction}\label{section:intro}

%\textcolor{blue}{Cambios en azul!}

Supermassive black holes (SMBHs) reside at the centres of most galaxies, and hence characterizing them is crucial for understanding galaxy evolution and the interplay between SMBHs and their hosts. One key aspect of SMBH research involves studying the demographics of SMBHs, including their mass distribution, occurrence rates, and correlations with host galaxy properties. This requires robust methods for identifying and classifying SMBHs across a wide range of environments. 

A powerful tool in this pursuit is the study of their variability properties. Active galactic nuclei (AGN), powered by accretion onto SMBHs, show significant variability in their luminosity across different wavelengths. Optical variability, in particular, offers the opportunity to study various physical processes of the accretion disc and surrounding regions \textcolor{blue}{\citep[e.g. reviews by][and references therein]{Cackett2021, Paolillo2025}}. In addition, the ubiquity of optical variability in AGN provides a robust and reliable method for their identification through optical monitoring campaigns as reconginzed early on by \citet[e.g.][]{Trevese1989}. For example, repeated photometric observations of a given region of the sky performed by the Sloan Digital Sky Survey \citep[SDSS;][]{York00} were used to select AGN and quasars using different variability strategies \citep[e.g.,][]{Schmidt10,PalanqueDelabrouille2016,Butler2011, Myers15}. \citet{Peters15} later demonstrated that a combination of variability and colour information improved the selection of AGN in SDSS, and have subsequently been exported to other data sets \citep[e.g.][]{DeCicco15,DeCicco2021, Tie17,Sanchez-Saez19,Sanchez-Saez2021,Sanchez-Saez2023}. 

Dedicated optical monitoring campaigns such as the Palomar Transient Factory \citep{Rau2009}, Pan-STARRS \citep{Chambers2019} and more recently the Zwicky Transient Facility \citep[ZTF;][]{Masci19} have revolutionized the identification and characterization of AGN by producing large datasets of light curves with sensitive photometry, consistent cadence, long durations and covering large areas of the sky. The temporal behaviour of these optical light curves can efficiently distinguish AGN from other sources and quantify their intrinsic variability properties. However, the quality of the optical light curves is critical for reliable variability analyses. Large variations in cadence, total length, depth, and photometric accuracy significantly impact our ability to detect and characterize variability signals.

ZTF currently offers the largest volume of publicly available light curves. These data have been used to select AGN, among others, by \citet{Sanchez-Saez2021} using the ZTF alert stream and by \citet[][hereafter \citetalias{Sanchez-Saez2023}]{Sanchez-Saez2023}, \citet{Healy24,Nakoneczny25}, using the full light curves available through the data releases (DRs). Although these classifiers are successful at identifying AGN, the alerts are limited to 5-$\sigma$ flux variations, which leave many known AGN undetected, while the DR light curves are built from PSF-photometry on the science images, which are not host-galaxy subtracted.  Nearby AGN in resolved galaxies appear as point sources overlaid on extended emission regions, the spatial blending of which varies dramatically due to the epoch-dependent PSF,
so neither DR aperture photometry nor PSF photometry can produce accurate nuclear fluxes or light curves. The best alternative is to perform aperture photometry on host-subtracted images, which are also provided by ZTF. We explore the benefit of this approach below. 

This paper is composed of three parts: First, we present the construction of the photometry and light curve from the reference-subtracted images in Sec.~\ref{section:part1}. We then quantify the improvement of these light curves compared to the aperture photometry on the science images provided in the ZTF DRs in Sec.~\ref{section:features_comparison}, as our goal is to produce more precise light curves and more accurate errors for robust and trustworthy estimates of the variability. These light curves have many uses, but herein we mainly explore how well they are able to separate variable AGN from non-variable galaxies. Second, we use these light curves to feed a random forest classifier and compare the outcome to an essentially identical classifier which uses ZTF light curves taken directly from the DRs, as described in Sec. \ref{section:classification}. 
Finally, we compare the samples of AGN candidates obtained by this new classifier to AGN candidates obtained by different methods, namely colour selections and X-ray detections, in Sec.~\ref{section:sample_comparison}.  

\section{Construction of forced, aperture photometry light curves from ZTF difference images, DI-Ap}\label{section:part1}

Our goal is to produce light curves extracted from the ZTF \emph{difference} images, forcing the photometry on the centres of all sources detected in the \emph{science} images of the same field, using only observations that match a set of quality criteria. For this purpose, we download metadata files from IRSA\footnote{i.e.\ https://irsa.ipac.caltech.edu/ibe/search/ztf/products/sci? WHERE=field=\$i\&COLUMNS=infobits,maglimit,seei
ng,ra,dec,field,ccdid,qid,filefracday,airmass,obsjd,filtercode\&ct=csv, where \$i stands for the field number.} to select only observations performed under good conditions by requiring ZTF labels {\tt infobits}\,$=$\,0, {\tt maglimit}\,$>$\,20, {\tt seeing}\,$<$\,4$"$. Of these, we further select only the first observation in a given night, requiring that the difference in MJD for the same field/CCD/quadrant/filter combination differ by more than 0.5 days; this avoids keeping epochs from few-day intensive monitoring campaigns, which we find produce less homogeneous light curves. A key interest is detecting AGN in an area that will be covered by the upcoming 4MOST telescope \citep{DeJong2019}, located in the Southern hemisphere, so we limited the download to $-29^\circ <$ dec $< +15^\circ$, where $-29^\circ$ is the southern limit of ZTF, and to Galactic latitudes $|b|\gtrsim20^\circ$ above and below the Galactic plane. We limited the analysis to this region due to computer capacity constraints.  The Galactic plane region in particular is out of the scope for this project due to the exponentially increasing number of sources involved. Moreover in the Galactic plane the extreme ratio of Galactic to extragalactic sources would cause significant contamination even for a good classifier. We study the effect of Galactic latitude on the density of extragalactic sources in Fig. \ref{fig:fractions_lat}.  

The observations were downloaded in September 2022, and thus our light curves only include relevant data publicly released up to that date (i.e., DR13). For each valid epoch, we downloaded the associated difference image and the catalogue of detected sources produced by ZTF in the corresponding science image. Additionally, we downloaded the photometric catalogues of the reference images, provided by ZTF DR13, for all the selected fields, to construct "total" fluxes for each epoch, considering that we will be measuring fluxes from the \emph {difference} (i.e. science minus reference) images.\footnote{Here, "total" flux refers to the values obtained from DR catalogues on stacked images, and hence is dependent on the average seeing for extended+nuclear sources. While it provides a relative anchor point to construct approximate total flux light curves, which is fine for our purposes, we caution that systematic offsets may exist when comparing to light curves constructed from different surveys or even other ZTF DRs.}

\subsection{Photometry}\label{section:ZTF}

The ZTF difference images available from the DRs are constructed by transforming a DR-specific reference image to each science image, to match the sky coordinates and PSF, and then subtracting the transformed reference from the science frame \citep{Masci19}. In this way, difference images and science images have the same geometry. We used the catalogue of objects detected in each science image as the detection catalogue for running the photometry package SExtractor \citep{Bertin1996} in dual mode, and used the corresponding difference image for the measurements. In this way, a flux measurement is obtained for each object detected in the science image of a given epoch, regardless of its signal in the corresponding difference image. We used a fixed aperture of 4$"$ to ensure $>$50\% coverage of the PSF in all epochs and incorporated a PSF correction factor in the calibration stage, as described below. All the measurements and following analyses are carried out for each quadrant independently. 

The total flux for each source is calculated adding the flux measured in the difference image to the flux in the reference image, as
\begin{equation}
    F_{\rm tot} = F_{\rm ref, tot} + F_{\rm diff}  
\end{equation}
where 
\begin{equation}
F_{\rm ref, tot} = 10.0^{-0.4*(m_{\rm ref, tot}-ZP_{\rm diff})}
\end{equation}
is the total flux in the reference image converted using the magnitude zero point of the difference image ($ZP_{\rm diff}$), to obtain instrumental fluxes in the same scale as the flux measured in the difference image, $F_{\rm diff}$. The quantity $m_{\rm ref, tot}$ is the magnitude tabulated in the reference image catalogue, measured with a fixed 4$"$ aperture. The total magnitude of each source and epoch is obtained from $F_{\rm tot}$ as
\begin{equation}
m_{\rm tot}=-2.5 \log{F_{\rm tot}} - ZP_{\rm diff}.
\end{equation}

\subsection{Calibration, errors, and light curve construction}
It is well established that the publicly available calibrated ZTF images and related photometry have typical uncorrected systematic photometric uncertainties at the $\approx$0.1 mag level,\footnote{e.g., http://nesssi.cacr.caltech.edu/ZTF/Web/Calib.html} which is potentially comparable to the typical variations we seek to detect from many AGN. Portions of these systematics arise from variations between fields, CCDs, quadrants, and even within quadrants due to a variety of factors, and have at least partially been attributed to the use of relatively rare, bright calibration stars. Using fainter calibration stars, additional improvements can be made.\footnote{e.g., http://nesssi.cacr.caltech.edu/ZTF/Web/Zuber.html}. Since our goal is to produce accurate light curves,
we aim to minimize the epoch-to-epoch variations of constant sources and obtain cleaner measures of the variability of variable sources. To this end, we adopt the following calibration procedure.

We use the photometric catalogues provided in the ZTF DRs obtained from the reference image of each quadrant to calibrate each epoch. In brief, for each epoch we match by coordinates keeping the nearest neighbour within $1\farcs.5$, the sources in the reference catalogue to sources in our photometric catalogue, keeping only point sources with well measured photometry in both catalogs, dim enough to not be saturated and bright enough to have small errors, while keeping as large a range of magnitudes as possible. With this in mind, we selected objects that had {\tt classtar}\,$>$\,0.7 in the reference catalogue, limited to the magnitude range 12\,$<$\,$g$\,$<$\,19.5 and that have magnitude errors less than 0.3 in both the reference catalogue provided by ZTF and in our SExtractor catalogue. If the number of remaining sources, hereafter 'calibration objects', is less than 15, the epoch is not calibrated and is not included in the light curve construction. 

We then calculate the magnitude difference for each calibration object, $m_{\rm diff}=m_{\rm tot}-m_{\rm ref}$, and fitted a first order polynomial to the resulting scatter plot of $m_{\rm diff}$ vs. $m_{\rm tot}$. The points that deviate from the fit by more than 5 times the root-mean-square of all the deviations are discarded and the polynomial is refit. If less than 10 calibration objects remain, then the epoch is excluded from light curve construction. This polynomial fit to $m_{\rm diff}$ vs. $m_{tot}$ is then used to correct all the magnitudes in our SExtractor catalogue as $m_{\rm cal}=m_{tot}$\,$-$\,polyfit$(m_{tot})$. This calibration is carried out for each quadrant independently. Lightcurves for a given object coming from different field/ccd/quadrant combinations are not combined.  

The errors on the magnitude produced by SExtractor from the difference images are severely underestimated for bright sources, where the Poisson noise of the source flux dominates over the uncertainties in the background subtraction. This happens because the difference images retain the Poisson noise of the science image but lose most or all of the flux, so the calculation of the Poisson noise is flawed. To remedy this problem, we measure the scatter of the magnitude differences $m_{\rm diff}$ for the previously mentioned calibration objects in bins of magnitude. We then interpolate the resulting rms vs median magnitude of the bins to predict an error as a function of magnitude for all the sources in our SExtractor catalogue. Finally, the code compares this error to the original error from SExtractor and retains the largest of the two. This is done to maintain the SExtractor errors for dim sources, where the magnitude uncertainty is dominated by the background. This correction and error estimation was done independently for each quadrant and epoch.

We then select the objects for which we will create light curves out of the calibrated catalogues described above. This selection was based on the ZTF-produced photometric catalogues on the reference images. These catalogues were filtered to include only objects with flag=0, to select well-detected, isolated objects, and magnitudes brighter than 20.5 (using the MAG-BEST and magnitude zero point columns) to discard the large number of dimmer objects for which the observational noise is typically larger than the variations of normal AGN. We assigned object identification codes (ID) for all objects in these filtered lists, incorporating field, filter, CCD, and quadrant number, along with a sequential source number. We note that a given object that was observed in several fields will have multiple object IDs. All the calibration and light curve construction is done for each quadrant independently, so an object with multiple IDs will also have multiple light curves.   

The light curves were built by cross-matching by coordinates the objects on these lists, to the detections across different epochs of the corresponding quadrants. The cross matching was made by selecting the nearest match, within $1.\farcs 5$.
Finally, all object IDs for which light curve construction was attempted were consolidated into a single ZTF-ID list containing over 42 million entries. 

\section{Comparison of variability features of DI-Ap light curves to DR11-psf  light curves}
\label{section:features_comparison}
The production of accurate light curves and errors can be particularly difficult for problematic objects such as spatially-resolved galaxies. In such objects, PSF photometry performed on science (not reference-subtracted) images can be inaccurate and the errors difficult to model. The latter is the method used to produce the light curves distributed in the ZTF DRs. Thus, it is instructive to compare some key variability features assessed for light curves from our DI-Ap vs. DR11-psf.For this purpose, we select objects across various classes identified in the labelled set used in \citetalias{Sanchez-Saez2023}, who generated variability features from DR11-psf  light curves. 

We cross-matched our ZTF-ID list to the  objects in the labelled set (LS) of \citetalias{Sanchez-Saez2023}, finding 451.327 matches, dominated by non-variable stars.
%We cross-matched our ZTF-ID list to the  objects in the labelled set of \citetalias{Sanchez-Saez2023} that lie between $-29^{\circ}$\,$<$\,dec\,$<$\,15$^{\circ}$ and $|b|>20^{\circ}$, finding 451,327 matches, dominated by non-variable stars.
%
Below we review the performance for non-variable stars, non-variable galaxies (i.e. low-redshift, mostly resolved galaxies expected to show no real variations), low-z AGN (i.e. galaxies with $z<0.5$ with identified nuclear activity, that should present real flux variations) and mid-z ($0.5<z<3$) AGN which should also present real fluctuations and are generally not spatially resolved in ZTF images.   

\subsection{Standard deviations}

\begin{figure*}
    \centering
    \includegraphics[width=0.43\textwidth]{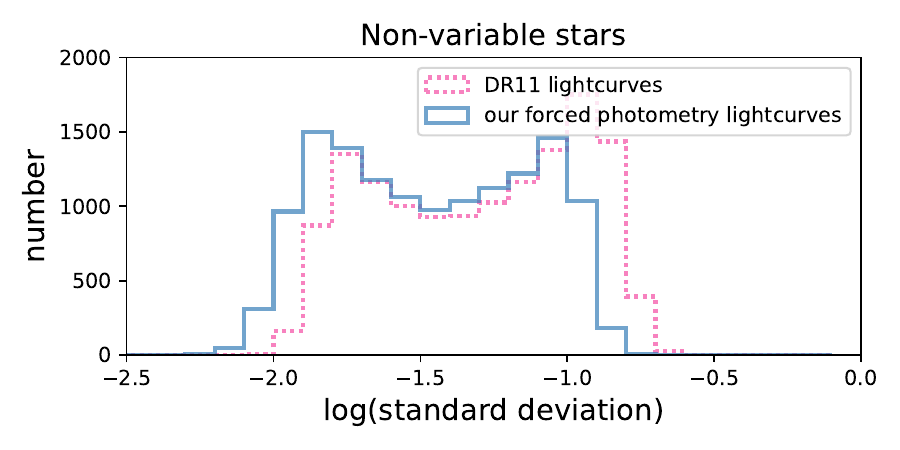}
     \includegraphics[width=0.43\textwidth]{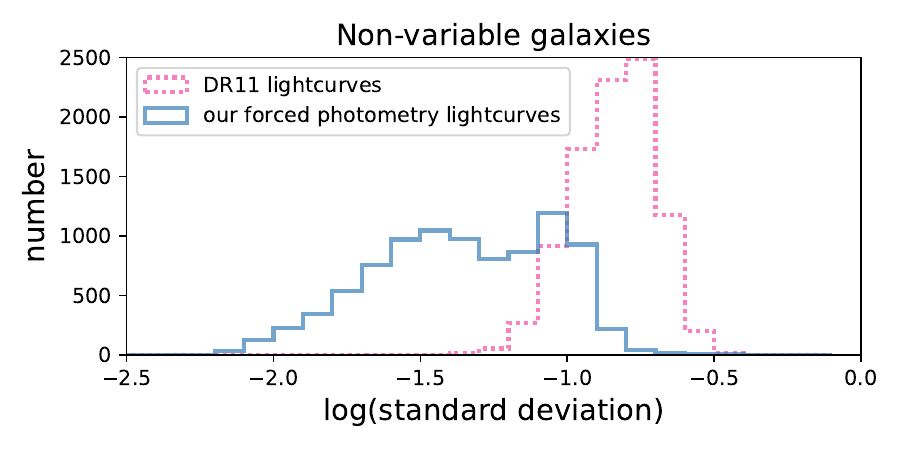}
    \includegraphics[width=0.43\textwidth]{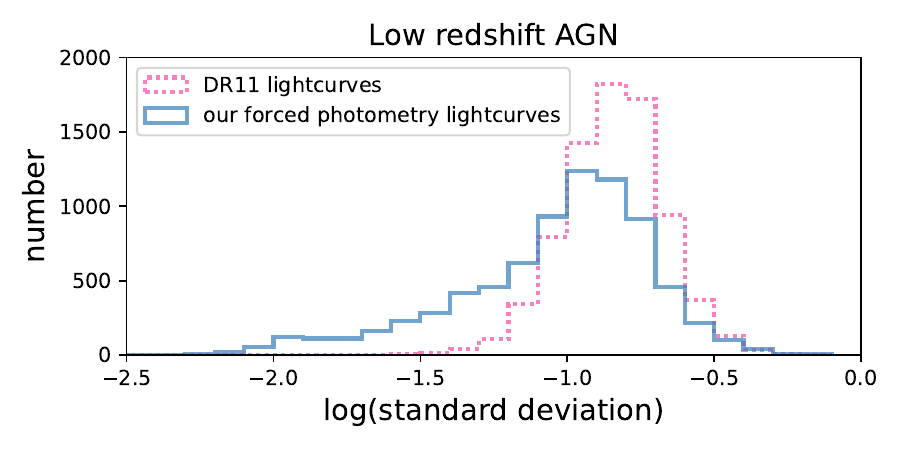}
     \includegraphics[width=0.43\textwidth]{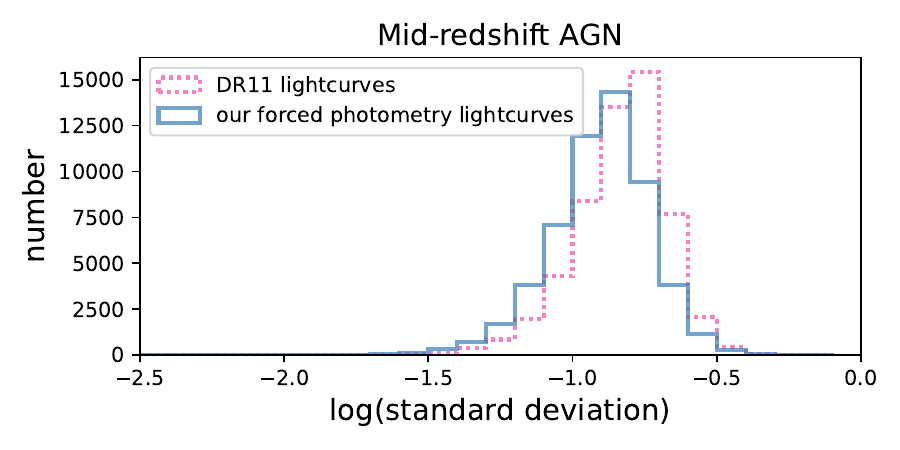}

   \caption{Distribution of the standard deviation of the light curves obtained from our custom-made photometry (blue, solid line) and directly from the ZTF data release DR11-psf light curves (pink, dashed line) for four types of objects in the labelled set: non-variable stars, non-variable galaxies, and low redshift ($z<0.5$) AGN and mid-redshift AGN ($0.5<z<3$). The standard deviation is calculated from the magnitudes.}
    \label{fig:std}
\end{figure*}

The standard deviation (hereafter {\tt stdev}) of the light curves is a simple measure of the total variations, integrated over the timescales sampled, and is independent of the estimation of the errors since the error bars do not enter its calculation. Ideally, this value would be 0 for non-variable objects. The observational noise, however, will add some variations to the flux and the measured standard deviations will result in positive values. We aim to obtain the most precise light curves, which should be reflected in lower values of the standard deviation i.e. flatter lightcurves, where the contribution of the observational noise to the scatter and therefore also to the {\tt stdev} is minimized. In Fig. \ref{fig:std} we show the distribution of this statistic for the light curves from DR11-psf and our DI-Ap, separately for each class of object. 

For non-variable stars, the {\tt stdev} distributions are similar, with a $\approx$0.1 dex shift to lower values for the DI-Ap, indicating that for this class of objects both types of light curves are equally flat (i.e. have similar values of {\tt stdev}). This plot reflects the expectation for the variations produced by observational noise in the ideal, point-source case, which is dominated by the background component at high magnitudes and by the Poisson component at low ones, producing the observed bi-modal distribution. 

For the non-variable galaxies, we find that the DI-Ap light curves have a similar bimodal {\tt stdev} distribution as for the non-variable stars while the DR light curves pile up at relatively high {\tt stdev} values. We note that the statistics were calculated for the same objects for the DR and DI-Ap light curves, so this stark difference in variability amplitudes is a result of the methods used for the photometry and calibration. 

The low-redshift AGN share similar distributions in distance, size and flux with the non-variable galaxies, but we anticipate that the AGN labelled set should show stronger signs of variability. Indeed we observe a marked shift to higher {\tt stdev}s among the vast majority of AGN for the DI-Ap  light curves. The DR11-psf light curves, on the other hand, show nearly identical {\tt stdev} distributions for both non-active and active nearby galaxies. Both distributions lie 0.1-0.2 dex above the non-variable star distributions, which might be misinterpreted as (spurious) variability.  The DI-Ap  light curves can distinguish between these two populations much better than the DR11-psf light curves. 

Finally, the mid-redshift AGN show similar {\tt stdev} values in the DI-Ap  light curves as in the DR light curves. This proves that the DI-Ap  does not simply lead to lower {\tt stdev}s--- intrinsically variable objects do show larger {\tt stdev}s.The similarity in both distributions is probably a consequence of the point source appearance of the mid-z AGN in the ZTF images, which reduces the problems of the psf-photometry on the science images. However, a small difference in variability amplitude remains. 

\subsection{Pvar}
 The significance of the variations is computed by comparing the measured variability with the variations expected from photometric errors alone, which should be quantified by the error bars on the light curve fluxes. Assuming that random variations in a constant source are accounted for in the error bars, the significance of the variations will be low, while if errors are underestimated then non-variable objects would be identified as significantly variable. This comparison is often quantified with {\tt Pvar} \citep{McLaughlin96}, which is a standard statistic for the probability that the measured variance in the light curve is intrinsic. Ideally, non-variable objects should have {\tt Pvar}$\sim0$. On the other hand, if intrinsic variations are detected, as expected for example for AGN light curves, then {\tt Pvar} should tend to 1. We note that a correct value of {\tt Pvar} relies not only on the errors being accurate but also on the assumption that the flux errors are Gaussian and on a $\chi^2$ distribution of the variances, which are probably not the case in these lightcurves. Therefore the interpretation of {\tt Pvar} on its own should consider these caveats. In our case we only use this value to compare the general accuracy of the error bars in both types of light curves and for different types of sources.

 \begin{table}[htpb]

  \begin{center}
    \caption{Percentage of objects with {\tt Pvar}>0.9 by class and light curve type.}
     \label{tab:Pvar}
    \begin{tabular}{l|c|c}
    &\multicolumn{2}{c}{\% with {\tt Pvar}>0.95}\\
    \hline
    Class&DR11&our phot.\\
    \hline
    Non-variable star& 74&7\\
    Non-variable galaxy&98&7\\
    low-z AGN&97&76\\
    mid-z AGN&96&88\\
    \end{tabular}
         \end{center}
\end{table} 
 
 Table \ref{tab:Pvar} compares the distributions of {\tt Pvar} calculated from our DI-Ap  and DR11-psf light curves for non-variable stars, non-variable galaxies and low-redshift and mid-redshift AGN in the labelled set. For the DI-Ap light curves, {\tt Pvar} concentrates toward low values for the non-variable classes, almost independently of the source morphology. This fact is encouraging because the non-variable galaxies in the labelled set correspond to low redshift objects, which are often resolved in the ZTF images, and the light curves of resolved sources are more likely to contain photometric errors which are not accounted for in the error bars. The low redshift AGN in the labelled set have by definition $z<0.5$ and are therefore also often resolved in the ZTF images. Their {\tt Pvar} distribution, however, differs strongly from the non-variable galaxies set, concentrating towards {\tt Pvar}=1. Therefore, the DI-Ap  light curves and associated error bars are sufficiently accurate to produce low {\tt Pvar} values for non-variable objects and high {\tt Pvar} values for variable objects, clearly distinguishing local active and non-active galaxies. The full distributions of {\tt Pvar} are shown in Appendix \ref{app:Pvar}.
 In contrast, the DR11-psf light curves produce nearly universal, significant variations ({\tt Pvar}$\ge$0.95) for all classes, indicating that the error bars are underestimated. 
 Although it is still possible to distinguish low-redshift AGN from non-variable galaxies using the DR11-psf light curves, as was shown by \citetalias{Sanchez-Saez2023}, the classification must rely on other variability and colour features. 
 
 This stark deviation from {\tt Pvar}$\sim$0 is an important caveat that must be considered when using the ZTF DR light curves to find variable objects with standard variability statistics, i.e., low redshift, intrinsically non-variable galaxies will appear variable and might be mistakenly classified as AGN. Similarly, a large fraction of non-variable stars might be incorrectly considered variable.

\subsection{DRW $\tau$}
The structure of the variations in AGN is often characterised by the Damped-Random-Walk (DRW) model, whose parameters $\sigma$ and $\tau$ correspond to the amplitude of long term fluctuations and the characteristic timescale of fluctuations, respectively. For AGN with black hole masses of $10^8-10^{10} M\odot$, this characteristic timescale is on the order of 100--1000 days \citep{Burke21,Tang2023,Arevalo2024}. In cases where the amplitude of variability is very small or indistinguishable from the noise, the fitted values of $\tau$ tend toward the sampling timescales, which is the only characteristic timescale of the light curve. Therefore, we expect values of $\tau\sim 1-3$ days for non-variable objects and of $\tau\sim 100-1000$ days for massive AGN, which correspond to the majority of the AGN in the labelled set. We caution the reader that the DRW might not be a good description of the AGN variability and is not the correct model for stellar pulsations, so the value of $\tau$ should not directly be interpreted as an accurate measure of a the characteristic timescales of the individual objects. We present them here because they do serve to compare the structure found in the lightcurves obtained from both methods and to distinguish long term fluctuations from uncorrelated, point-to-point noise. 

\begin{figure*}
    \centering
    \includegraphics[width=0.45\textwidth]{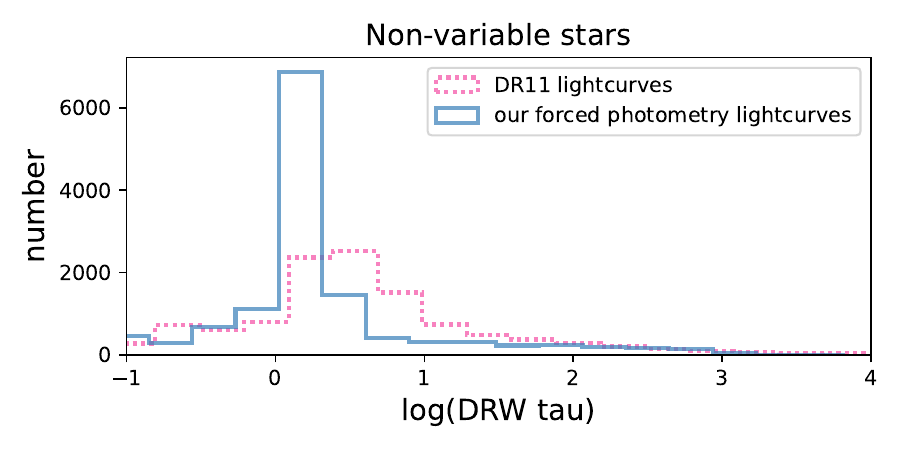}
     \includegraphics[width=0.45\textwidth]{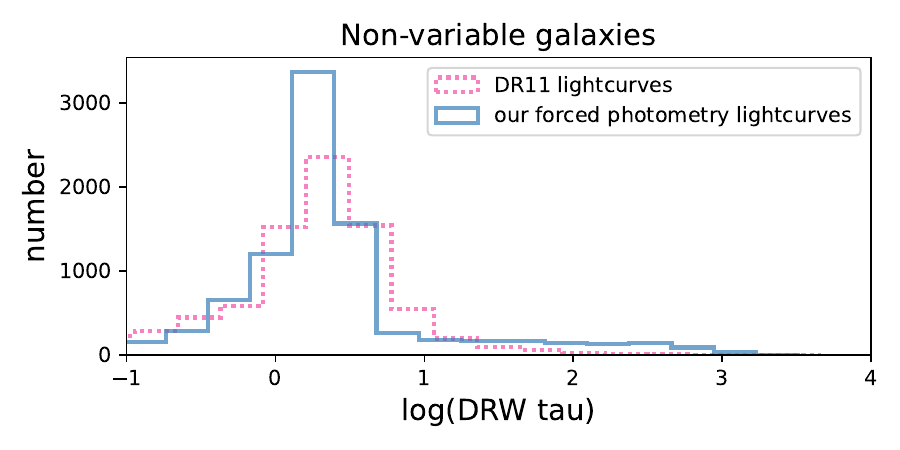}
    \includegraphics[width=0.45\textwidth]{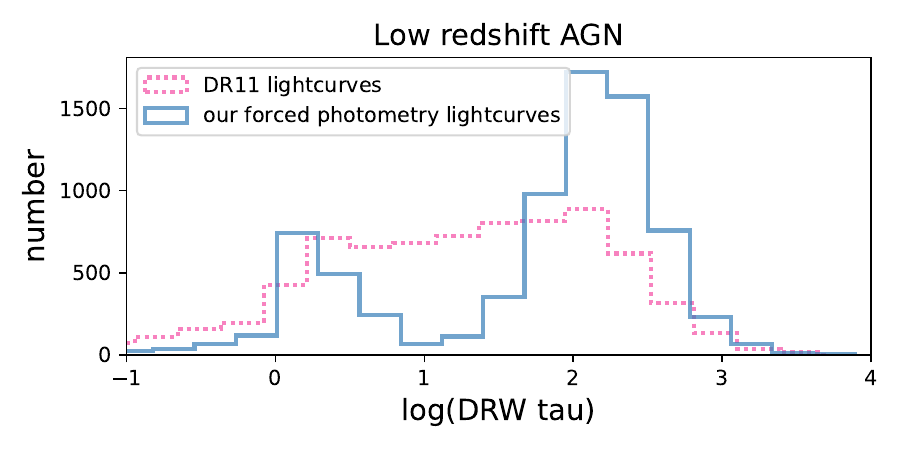}
     \includegraphics[width=0.45\textwidth]{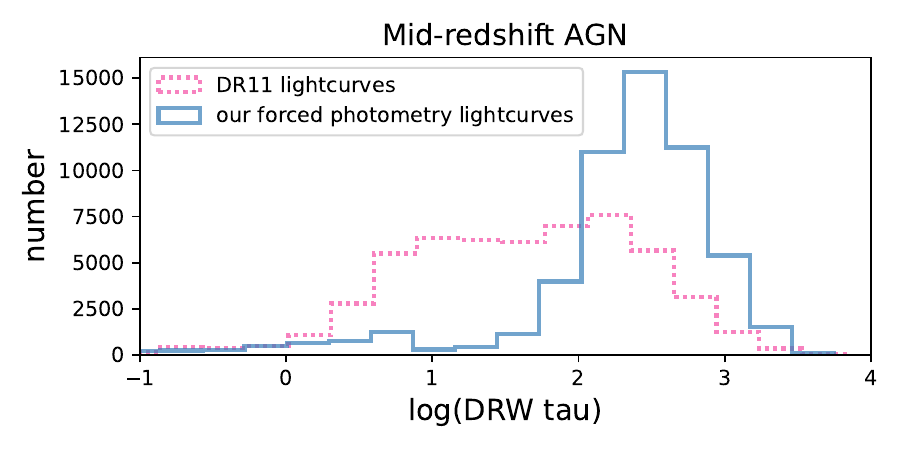}

   \caption{Distribution of DRW $\tau$ values associated with our DI-Ap  light curves (blue, solid line) and DR11-psf  light curves (pink, dashed line), respectively, for non variable stars, non-variable galaxies, low redshift ($z<0.5$) AGN and mid-redshift AGN ($0.5<z<3$) in the labelled set.}
    \label{fig:tau}
\end{figure*}

Figure \ref{fig:tau} shows the distributions of the DRW $\tau$ values fitted to the DR11-psf and DI-Ap  light curves for non-variable objects (top panels) and variable AGN (bottom panels). For both sets of light curves, the non-variable objects cluster toward $\tau\sim 1-3 $ days, with only slightly longer characteristic timescales for the DR11-psf light curves, which may reflect small spurious trends in the data. In the case of variable objects, the DI-Ap  produces values of $\tau$ either concentrated towards the sampling timescale or toward the expected timescale of 100--1000 days, with the latter peak being more populated for the mid-redshift AGN than for the low redshift AGN. The concentration at $\tau\sim 1-3 $ days is probably caused by AGN with little or no detected variability (i.e., according to {\tt Pvar} or the standard deviation), which are more numerous in the low redshift bin. In either case, the DR11-psf light curves result many times in intermediate values of $\tau\sim 3-100$ days, which are probably spurious. 
 
\section{Object classification using Random Forest algorithms }
\label{section:classification}
\subsection{Features}\label{section:features}

We computed variability features from the DI-Ap  light curves to help classify all available objects into different classes of astronomical sources. The method closely follows the classifier presented in \citetalias{Sanchez-Saez2023}, and thus we computed the same features used there (for a detailed description of the features see \citealt{Sanchez-Saez2021}), but also included the following additional features: Pan-STARRS $i-z$ colours; {\it Gaia} proper motion information; the Mexican-Hat \citep{Arevalo12} filtered variance at timescales of 45 and 450 days; the error on the excess variance; and a flux asymmetry estimator given by Asymmetry$=(N^+-N^-)/(N^++N^-)$, where $N^+$ is the number of light curve points above the mean flux and $N^-$ is the number of points below the mean.

\subsection{Model}
As mentioned above, the classification algorithm closely follows that presented in \citetalias{Sanchez-Saez2023} with the inclusion of a few new features and an identical classification taxonomy (see Fig. 1 of \citetalias{Sanchez-Saez2023}), where the sources are classified in a hierarchical fashion into 17 different classes (see Table \ref{table:scores} for the full list of classes).  Most relevant for this work are the AGN classes, where lowz-AGN corresponds to AGNs with $z\leq0.5$, midz-AGN to AGNs with $0.5<z\leq3$, highz-AGN to AGNs with $z>3$, and Blazars to beamed/jet-dominated AGNs, as well as the nonvar-galaxy (non-variable galaxy/extended source) and nonvar-star (non-variable star/point source) classes. These splits were motivated by the different observational properties of flux-limited AGN samples at different redshifts, considering shifts in colour and in characteristic timescales as a combination of different typical black hole masses and of cosmological time dilation. With this hierarchical approach, we have a total of 17 classes.

The LS needed to train our classification model was constructed using the master catalogue compiled by \citetalias{Sanchez-Saez2023}. This master catalogue includes several catalogues from the literature of known sources belonging to the classes listed in Table \ref{table:scores}. For the AGN classes, the relevant catalogues are the Million Quasars catalogue (MILLIQUAS catalogue v7.4c; \citealt{Flesch19}), the Roma-BZCAT Multi-Frequency catalogue of Blazars (ROMABZCAT; \citealt{Massaro15}), and a catalogue of Type 1 AGN from SDSS (Oh2015; \citealt{Oh15}). The master catalog presented in \citetalias{Sanchez-Saez2023} contains 1903799 sources, including all the classes listed in Table \ref{table:scores}. To construct our LS, we crossmatched the master catalogue with our ZTF-ID list, and we removed all the sources with an average magnitude $g>21.0$ or $g<13.5$. The number of sources per class included in the resulting LS is five times the number of sources shown in Table \ref{table:scores}.

We used 80\% of the LS as training set and 20\% as testing set, using the tool \texttt{train\_test\_split} available in \texttt{scikit-learn} \citep{Pedregosa12}, in a stratified fashion. 
Table \ref{table:scores} shows the high class imbalance present in our labelled set. Therefore, following the classification approaches of \cite{Sanchez-Saez2021} and \citetalias{Sanchez-Saez2023}, we used a balanced random forest (BRF; \citealt{Chen04}) in a hierarchical fashion (balanced hierarchical random forest; BHRF), using a local classifier per parent node approach \citep{Silla11}. The BHRF classifier used in this work has three levels. The first level (node\_init), separates the sources as nonvar-galaxy, nonvar-star, or variable. Then, all the sources classified as variable go to the second level (node\_variable), where they are separated as transient, stochastic, or periodic. Finally, in the third level, the transients are separated into SNIa/SN-other/CV-Nova (node\_transient), the stochastics into lowz-AGN/midz-AGN/highz-AGN/Blazar/YSO (node\_stochastic), and the periodics into LPV/RRL/CEP/EA/EB-EW/DSCT/Periodic-other (node\_periodic). Each node corresponds to a BRF, using the implementation of the \texttt{Imbalanced-learn} Python package \citep{imblearn}, with 300 estimators, a fraction of features to consider in each split (max features) of \texttt{sqrt}, a maximum depth of each tree (max depth) of \texttt{None}, a split criterion of \texttt{entropy}, and a class\_weight parameter of \texttt{balanced\_subsample}. The rest of the parameters were kept as the default values. 

For all the sources, the model provides probabilities and classifications of the node\_init, and only for the ones classified as variable, the model provides the probabilities of the node\_variable. The probabilities of the third level are provided depending on the class of each source in the node\_variable, namely, for transients, it provides the predictions of the node\_transient, for stochastic objects, the probabilities of the node\_stochastic, and for periodic sources, the probabilities of the node\_periodic. We did not calibrate the probabilities of the different nodes, but considering the similarities between the model presented here and the model used in \citetalias{Sanchez-Saez2023} (see Section 5.4 of \citetalias{Sanchez-Saez2023}), we expect that in general, the probabilities will underestimate the quality of the prediction in all the nodes, except for the node\_transient, for which the results are not optimal, as the light curves were constructed using the coordinates of the source in the template image, which is poorly suited for transients, as they can be off-nuclear.

\begin{table}[htpb!]
  \begin{center}
    \caption{\textbf{HBRF classifier metrics per class and macro-averaged scores.}}
    \label{table:scores}
    \begin{tabular}{ccccc} 
   
   \hline

Class & Precision & Recall & F1-score & \# sources \\

\hline

\hline
 
          SNIa   &    0.59   &   0.36  &    0.45  &     105 \\
      SN-other   &    0.17   &   0.48  &    0.26  &      23 \\ 
       CV/Nova   &    0.19   &   0.93  &    0.32  &      27 \\
      lowz-AGN   &    0.54   &   0.55  &    0.55  &    1952 \\
      midz-AGN   &    0.99   &   0.79  &    0.88  &   13717 \\
     highz-AGN   &    0.24   &   0.91  &    0.37  &     564 \\
        Blazar   &    0.15   &   0.69  &    0.25  &     106 \\
           YSO   &    0.83   &   0.56  &    0.67  &     177 \\
           LPV   &    0.94   &   0.90  &    0.92  &     202 \\
            EA   &    0.66   &   0.90  &    0.76  &     390 \\
         EB/EW   &    0.94   &   0.80  &    0.86  &    2335 \\
          DSCT   &    0.46   &   0.94  &    0.62  &      47 \\
           RRL   &    0.98   &   0.87  &    0.92  &    2458 \\
           CEP   &    0.07   &   0.60  &    0.12  &      25 \\
Periodic-other   &    0.25   &   0.60  &    0.35  &     178 \\
 nonvar-galaxy   &    0.79   &   0.98  &    0.87  &    2691 \\
   nonvar-star   &    1.00   &   1.00  &    1.00  &   63481 \\

  \hline

     macro avg  &     0.58   &   0.76   &   0.60   &  88478 \\

\hline

      accuracy  &           &        &      0.94  &   88478 \\

\hline
\hline

  \end{tabular}
 \end{center}
\end{table}

\begin{figure*}
  \centering
  \begin{minipage}{0.6\textwidth} 
    \centering
    \includegraphics[width=0.48\textwidth]{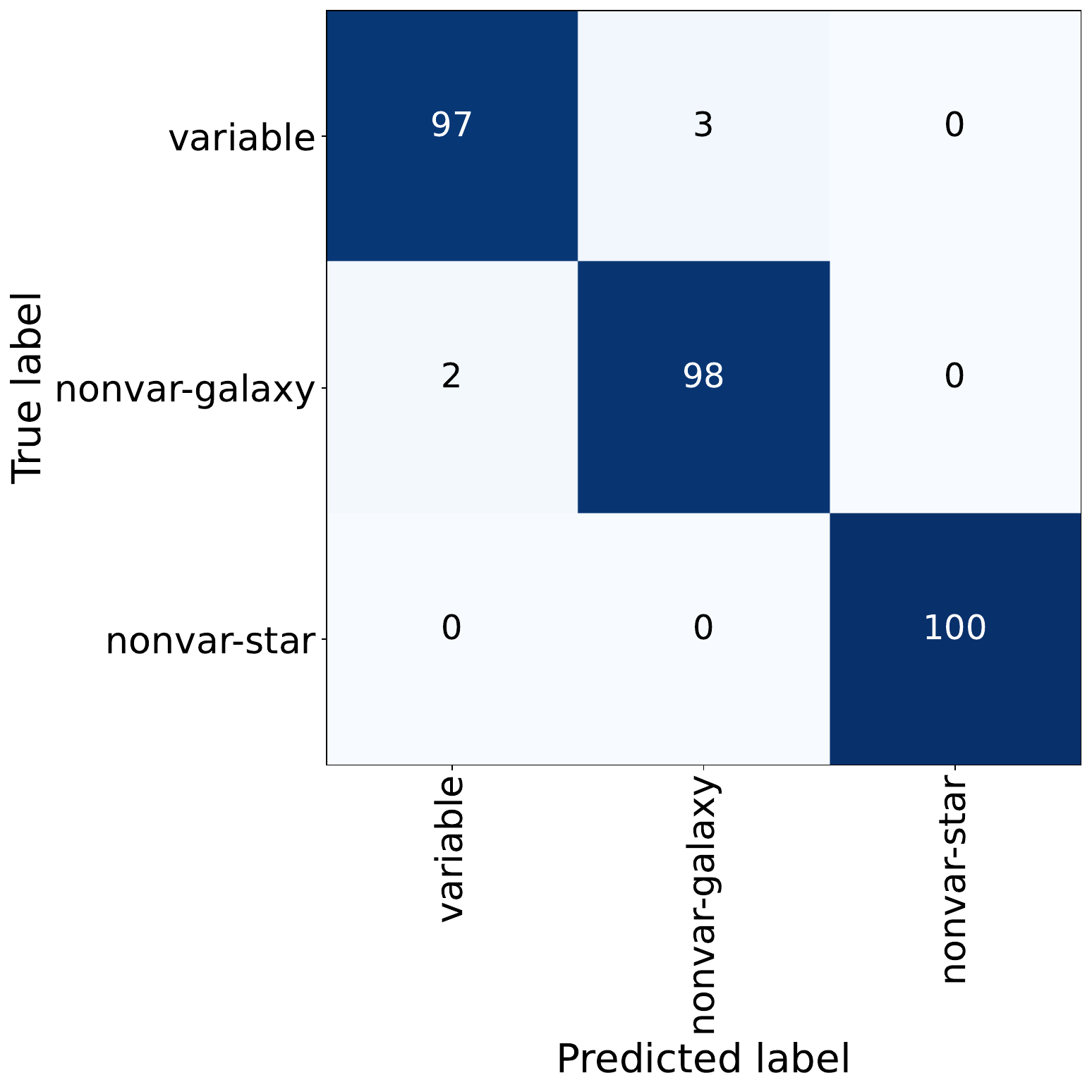}    
    \includegraphics[width=0.48\textwidth]{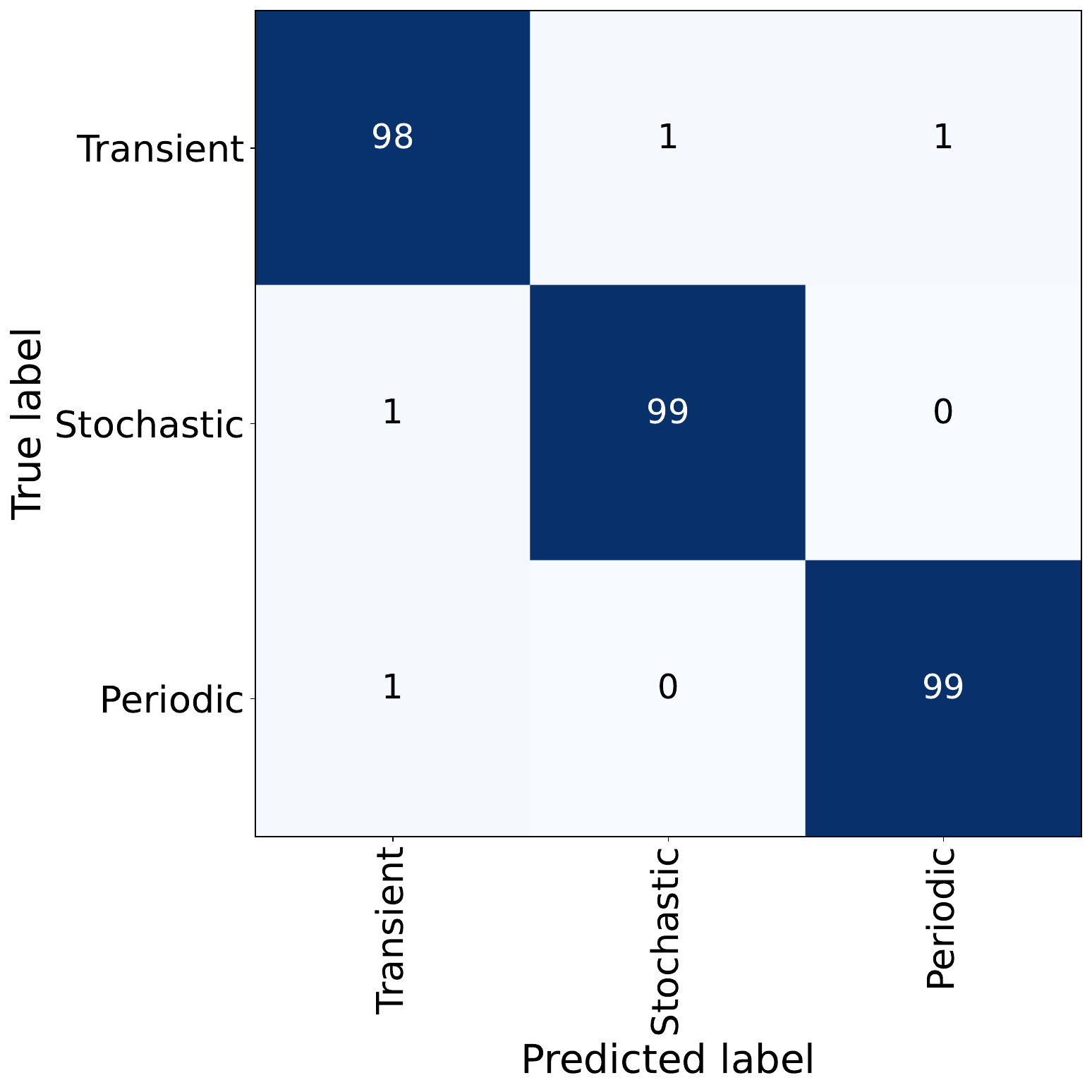}
  \end{minipage}%
  \hfill
  \begin{minipage}{0.40\textwidth} 
    \centering
    \includegraphics[width=\linewidth]{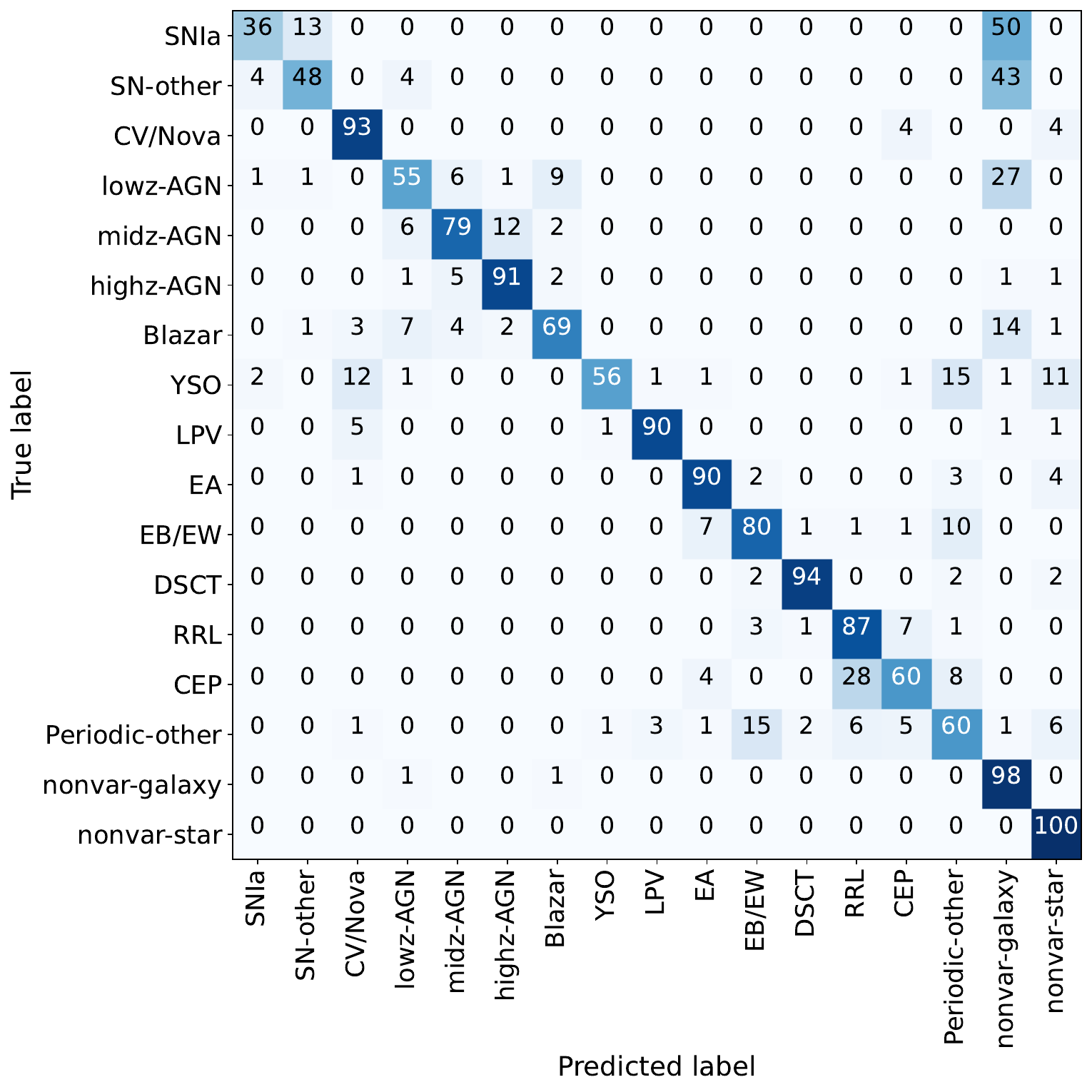}
  \end{minipage}
  
  \caption{Confusion matrices of the node\_init (top-left), node\_variable (bottom-left), and third level (right) obtained by using the HBRF in the testing set. The confusion matrices show the results as percentages, rounded to integer values, which are computed by dividing each row by the total number of objects with True labels.}
  \label{fig:confmat}
\end{figure*}

We evaluated the performance of our BHRF using the testing set. Table \ref{table:scores} shows the \texttt{classification\_report} method available in \texttt{scikit-learn}, which includes the precision, recall, and F1-score\footnote{For a detailed definition see Section 52 of \citetalias{Sanchez-Saez2023}.} for each of the 17 classes independently, and for the full testing set (macro-averaged scores). Moreover, Fig. \ref{fig:confmat} shows the confusion matrices of the node\_init (top-left panel), node\_variable (top-right panel), and of the third level (including the 17 classes; bottom panel).

From Table \ref{table:scores} and Fig. \ref{fig:confmat}, we can see that for the AGN classes, most of the confusion is among the AGN classes themselves or with the nonvar-galaxies. Grouping all the AGN classes as one class gives a precision, recall, and f1-score of 1.00, 0.96, 0.98, respectively. The lowz-AGNs and Blazars are the classes that the model tends to identify, for non-negligible fractions, as nonvar-galaxies. The same was observed in the results presented by \citetalias{Sanchez-Saez2023}; however, in the work presented here, the contamination is lower. Specifically, low-z AGN classified as non-variable galaxies drop from 36\% to 27\% and blazars classified as non-variable galaxies drop from 21\% to 14\% when replacing the DR11-psf lightcurves with our DI-Ap  light curves. This is expected, as now the light curves are less contaminated by the host galaxy flux, and are less sensitive to variations due to changes in the ZTF PSF.  
 
\subsection{Classification based on the DI-Ap  light curves}

\begin{figure}
    \centering
    \includegraphics[width=0.5\textwidth]{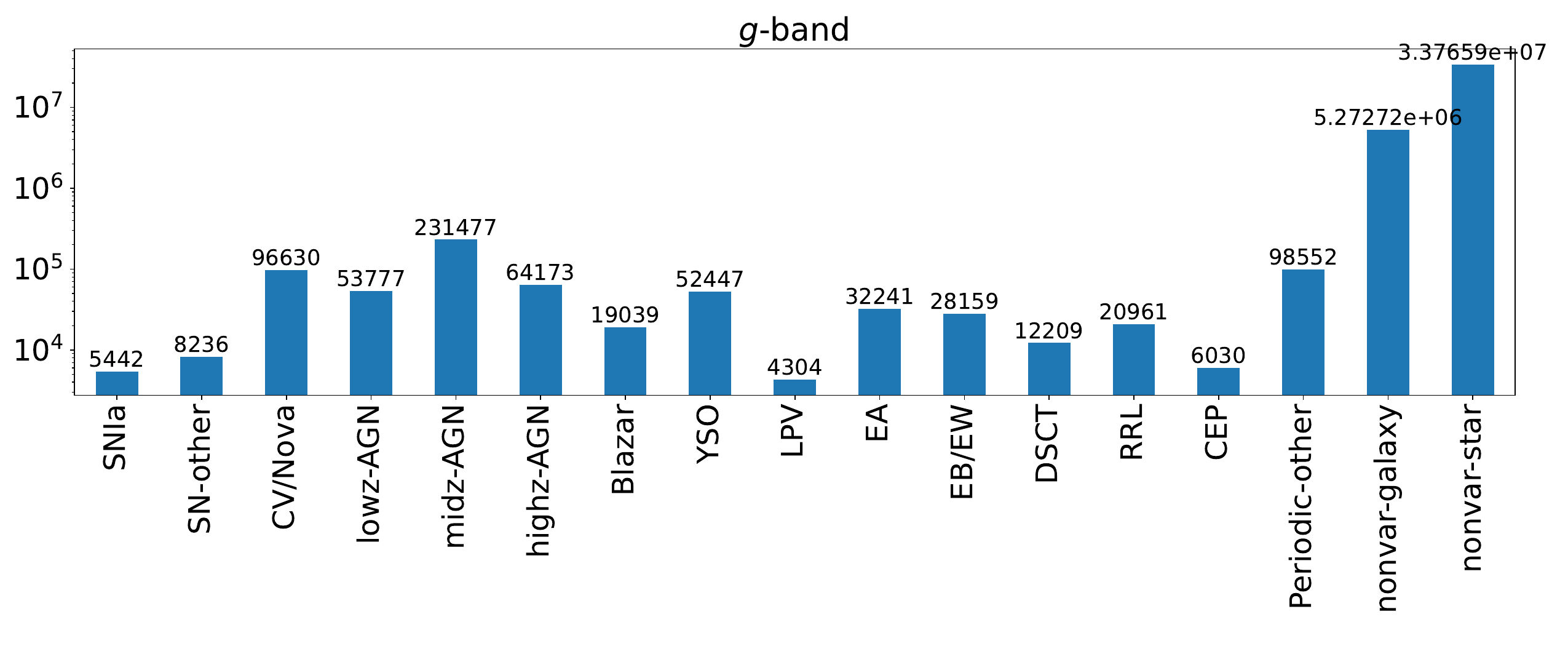}
     \includegraphics[width=0.5\textwidth]{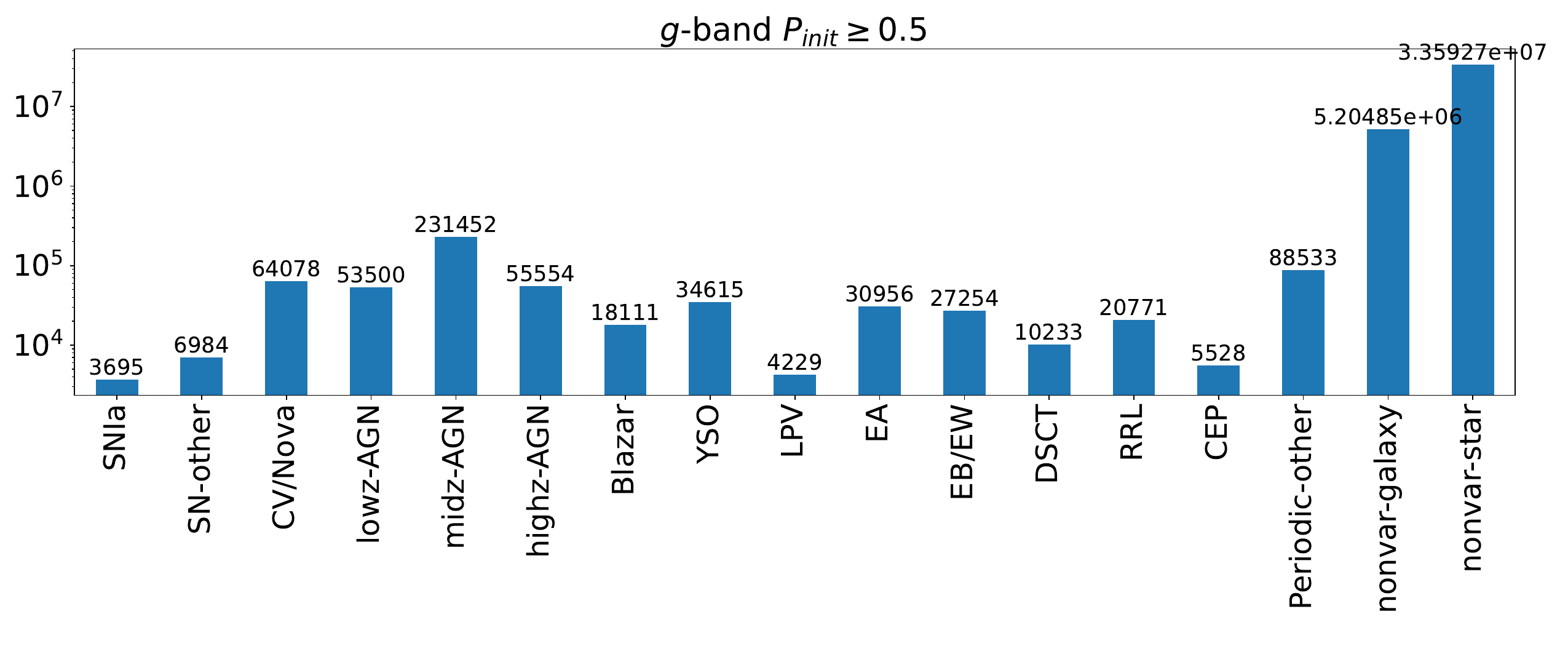}     

   \caption{Number of candidates per class for all the sources in the ZTF-ID list (top; 39,772,280 sources in total), and for the sources with a probability in the node\_init $P_{init}\geq0.7$ (bottom; 37,988,187). The number of sources per class is shown on top of each bar. }
    \label{fig:numclass}
\end{figure}

We used the HBRF model presented in the previous section to classify all the sources in the ZTF-ID list. In total, 39,772,280 sources were classified. The top panel of Fig. \ref{fig:numclass} shows the number of sources per class for the 17 final classes considered in this work, and the bottom panel shows the numbers when the sources are filtered by the probability of the node\_init ($P_{init}\geq0.5$), leaving a total of 37,988,187 objects. It can be seen that the majority classes, as expected, are nonvar-stars and nonvar-galaxies, followed by the midz-AGNs.

\begin{figure*}
    \centering
    \includegraphics[width=0.32\textwidth]{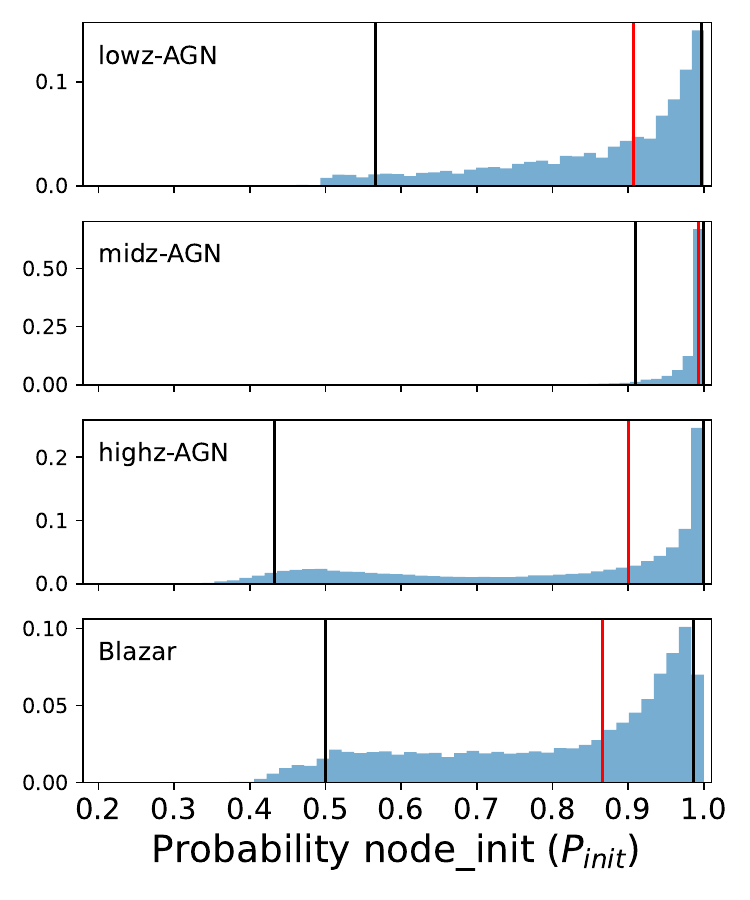}
    \includegraphics[width=0.32\textwidth]{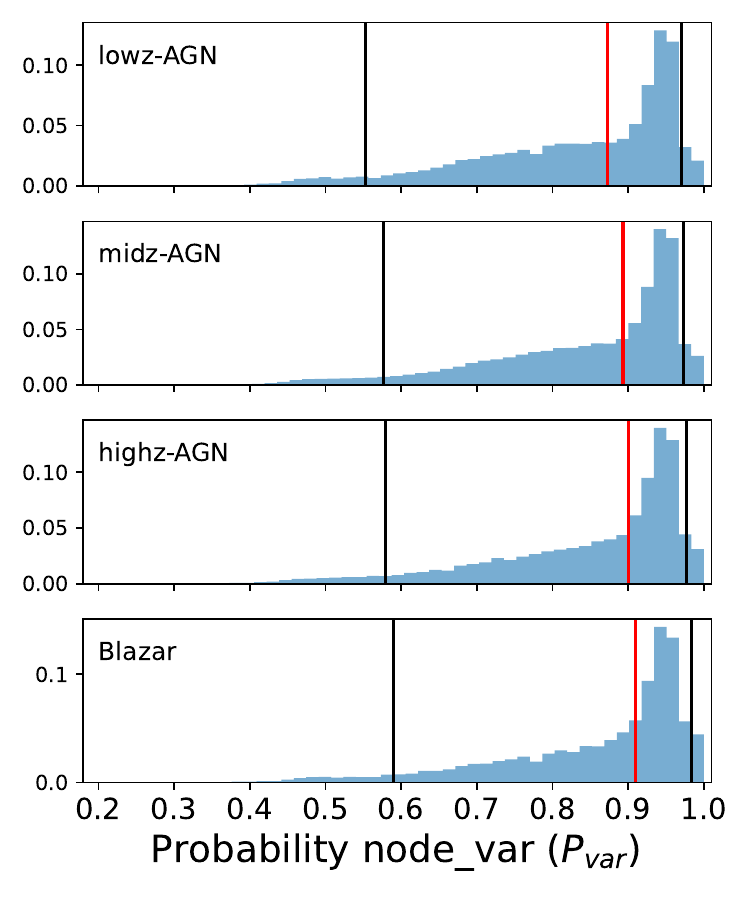} 
    \includegraphics[width=0.32\textwidth]{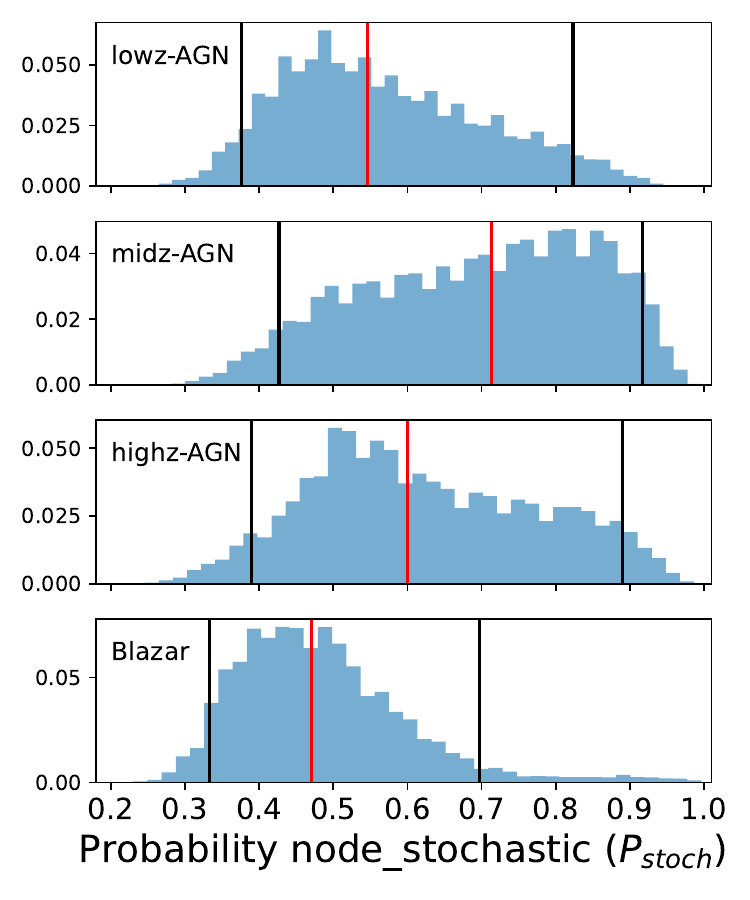} 

   \caption{Normalized probabilities of the node\_init (left), node\_variable (center), and node\_stochastic (right), for the four AGN classes. The red lines show the median probability for each class. The black lines show the 5th and 95th percentiles of the probabilities.}
    \label{fig:histprob}
\end{figure*}

In Fig. \ref{fig:histprob} we show the probability distributions of the nodes init, variable, and stochastic, for the four AGN classes. We can see that, in general, the probabilities of the node\_init and node\_var are very close to 1, this mean the model is pretty confident in that the sources selected as AGNs are variable and stochastic. The probabilities of the node\_stochastic show a broader distribution, but this is due to internal confusion between the four AGN classes, which is also seen in the confusion matrix presented in Fig. \ref{fig:confmat}, where it can be noted, for instance, that 12\% of midz-AGNs are misclassified as highz-AGNs, or 9\% of the lowz-AGNs are misclassified as Blazar. This confusion is expected, as the different AGN classes are just used to better identify sources at different redshifts, but do not represent an intrinsic distinct population of objects.

\subsection{Comparison with other ZTF-based classifications}
\label{section:comparative_classification}

\begin{figure*}
    \centering
 
     \includegraphics[width=0.49\textwidth]{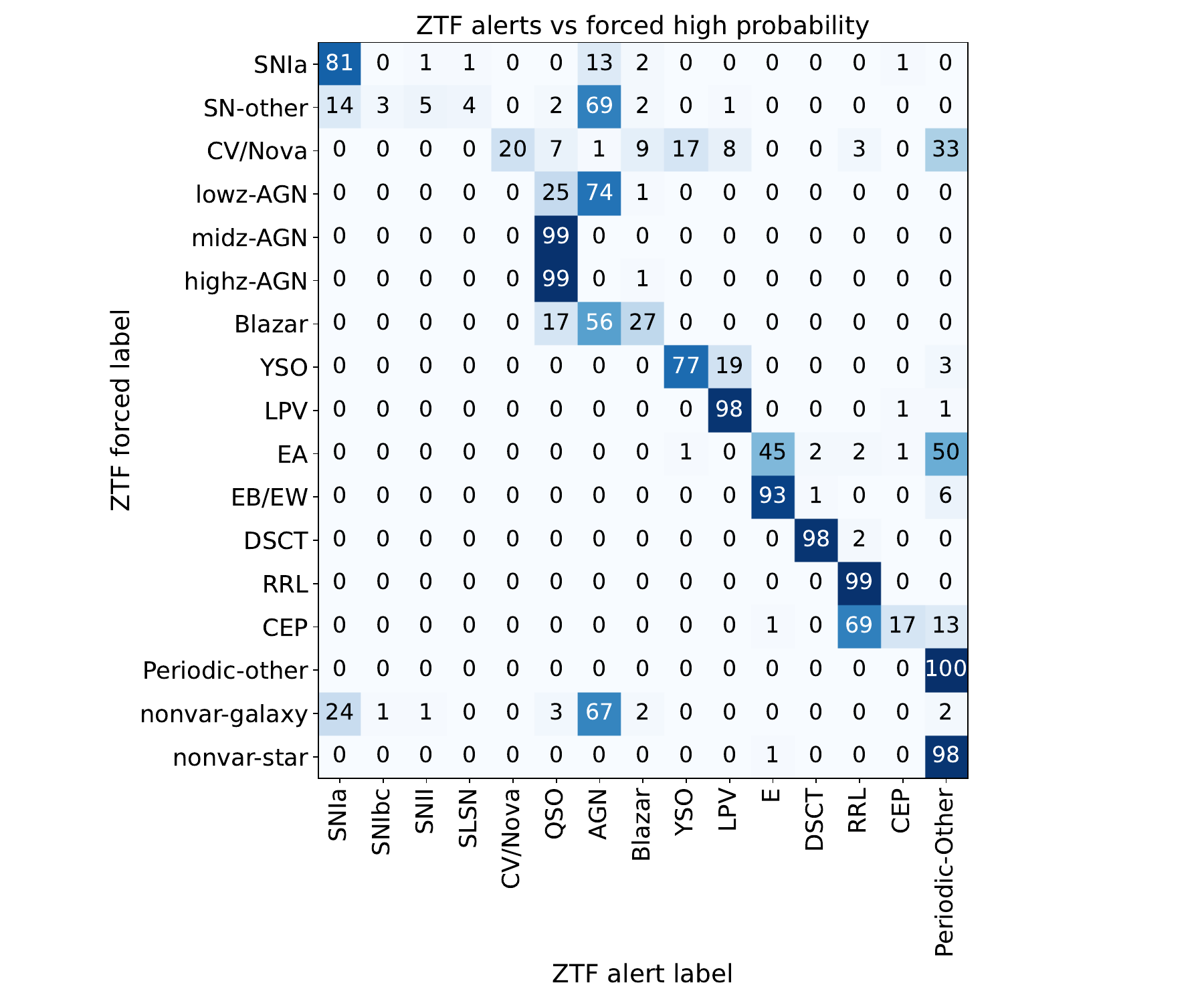 }    \includegraphics[width=0.49\textwidth]{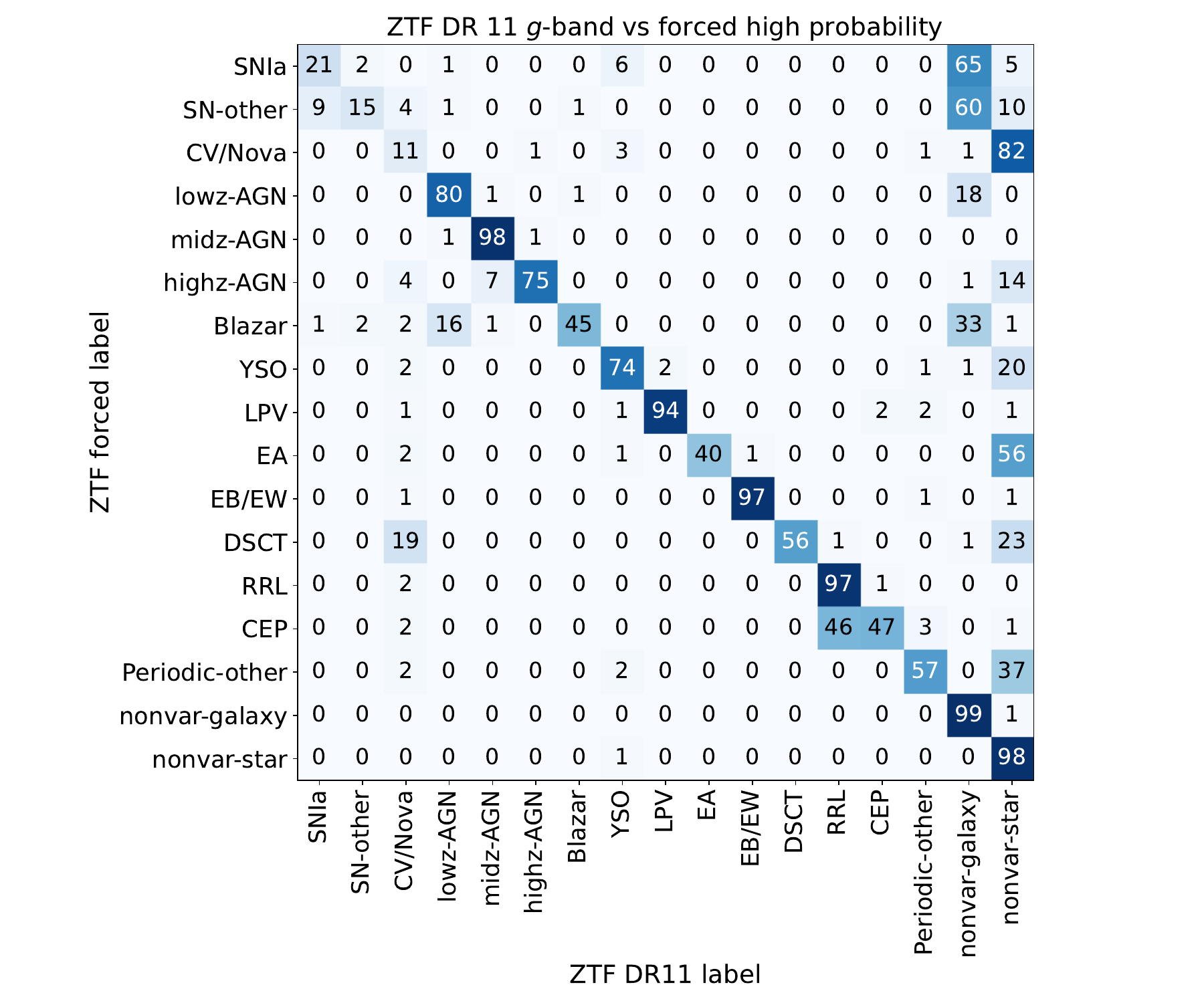  }
   
    \caption{Comparison between classifications from ZTF DI-Ap  vs. ZTF alerts from \cite{Sanchez-Saez2021} (left) and DR11-psf  from \citetalias{Sanchez-Saez2023} (right). We only include in these comparisons sources with probabilities larger than 0.5 in the final class. We divided each row by the number of objects per class with DI-Ap  labels (with $P\geq0.5$), which were included in the analysis, and we rounded these percentages to integer values.}
    \label{fig:comparison-otherztf}
\end{figure*}

The model used in this work inherits from the models used in \cite{Sanchez-Saez2021} (ZTF alerts) and \citetalias{Sanchez-Saez2023} (DR11-psf ). Therefore, here we compare the results of the classifications obtained using the ZTF alerts, with data collected between March 2018 and November 2022 by the ALeRCE broker \citep{Forster21}, and the classifications provided by \citetalias{Sanchez-Saez2023} for DR11-psf , with our DI-Ap  classifications. For this, we crossmatched our classifications (using our full sample, not just the labelled set) with the alerts and DR11-psf classifications (using their respective full samples as well), using a matching radius of 1\farcs5. This gave a total of 169,231 sources with both DI-Ap  and alerts classifications, and 32,835,087 sources with both DI-Ap  and DR11-psf classifications. Figure \ref{fig:comparison-otherztf} shows two comparison matrices, on the left for our ZTF DI-Ap classification versus the alerts classification, and on the right for our classification versus the classification of the DR11-psf  in the $g$-band. On both panels, we only include sources with probabilities larger than 0.5 in the final layer to remove potential spurious classifications, with 63,891 sources included in the DI-Ap vs alerts comparison and 32,198,401 sources in the DI-Ap vs DR11-psf comparison. Note that when comparing the DI-Ap  and the DR11-psf classifications, and keeping only the high probability candidates, only $\sim2\%$ of the targets are removed, while when comparing the alerts vs the DI-Ap  classification, using only the high probability sources, $\sim62\%$ of the targets are removed. This is partly due to the definition of the probabilities in the alerts model, which multiplies the probabilities of the top and bottom levels of the classifier, while the DR11-psf and the DI-Ap  models use a local classifier per parent node approach, which only provides the probability if the final node.

For the DI-Ap  versus alerts comparison, we can see that, in general, there is a good agreement between the two models, but as the alert classifier does not include the non-variable classes (by definition alerts only include epochs with a 5$\sigma$ detection in the difference images, thus only variables and transients are included in the alerts' model), it is not possible for the alerts' model to identify non-variable sources that appear in the ZTF alert stream. The presence of non-variable targets in the ZTF alert stream can happen for several reasons, including the presence of a transient in the template image or bad subtractions. This may explain the SN and AGN classifications in the alert stream that are classified as nonvar-galaxy in our model. 
The same happens for the periodic objects in the alert stream that are classified as nonvar-star by our model. More relevant for the AGN classes is the confusion between SN-other in the DI-Ap  and AGN in the alerts, which could imply that we are losing a small fraction of AGNs as transients. This is probably related to the lack of non-detection features in our model, which are included in the alerts model of ALeRCE and help to identify transient objects better, as well as the fact that our DI-Ap  light curves were not designed to deal with off-nuclear transients. However, we note that the classification of SLSN in the alert stream tends to be contaminated by AGNs showing $>$1--2 mag rise-fall activity. We also see internal confusion between the different AGN classes, but this is expected, considering that the model for the alerts uses a cut in luminosity and redshift, while our taxonomy considers a cut only in redshift. 

The right panel of Fig. \ref{fig:comparison-otherztf} compares the DI-Ap  and the DR11-psf model of \citetalias{Sanchez-Saez2023}. The results are in much better agreement compared to the alerts versus DI-Ap  comparison, due to the stronger model and taxonomy similarity. Importantly, we find that our DI-Ap  classifications recover a significant fraction of variable sources, particularly among AGN classes, but also for other transient and periodic classes, that were classified as non-variable in DR11-psf . This is expected, as the DI-Ap  light curves are less affected by systematic noise due to the DR11 PSF photometry and calibration, and, thus, are more sensitive to low-amplitude variations. 
This is particularly relevant for the identification of low-luminosity and host-dominated AGNs, whose variations could be hidden by the blending of host and AGN light in the DR11 PSF photometry. 

\subsection{Additional Cleaning}
\label{section:cleaning}
\begin{figure}
    \centering
    \includegraphics[width=0.48\textwidth]{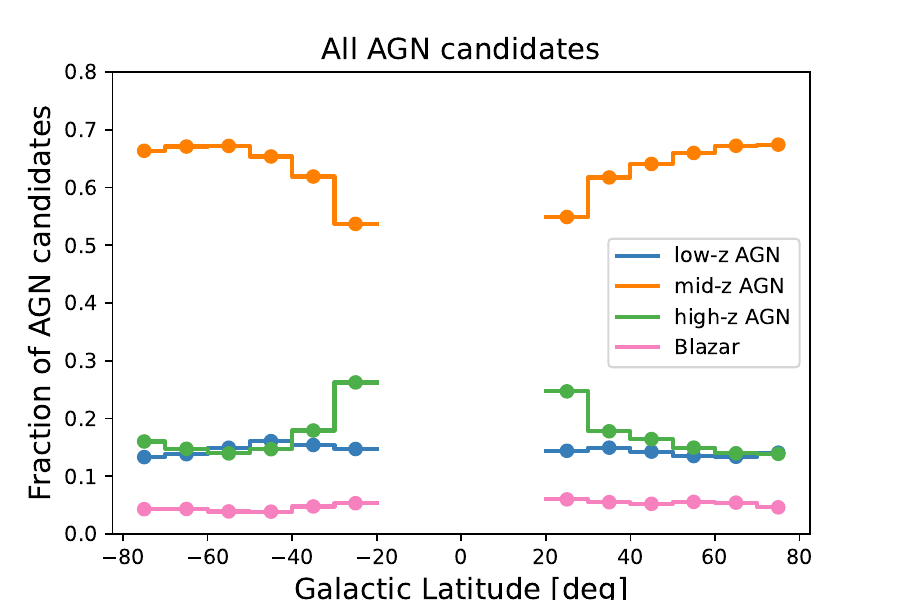  }
     \includegraphics[width=0.48\textwidth]{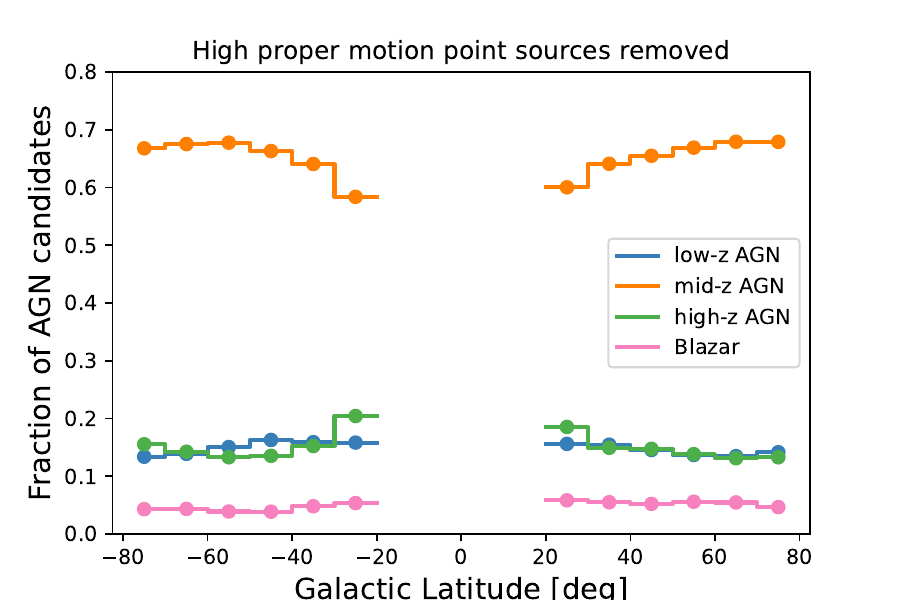  }
   
    \caption{AGN candidates per class divided by the total AGN candidates in a given bin of Galactic latitude. The top panel shows the distribution of all candidates, the bottom panel shows the distribution after removing point source candidates with significant proper motion (i.e. ps-score$>0.8$ and PM/PM error $>4$). 
    }
    \label{fig:fractions_lat}
\end{figure}

Special attention must be paid to AGN candidates which are point-like and exhibit significant proper motions according to {\it Gaia}, since these have a high probability of being Galactic sources misclassified as AGN. We note that extended objects can appear to have significant proper motions in {\it Gaia} due to errors in the {\it Gaia} photometry, so we do not remove these from the sample. If there is contamination from Galactic sources, these should concentrate toward the Galactic plane, while extragalactic sources should be independent of this variable. Thus, we investigate the dependence of different candidates populations as a function of Galactic latitude. The top panel in Fig. \ref{fig:fractions_lat} shows the fraction of AGN candidates classified as either Blazar, low-z, mid-z or high-z AGN in different bins in Galactic latitude. We note that, by construction, if one class is significantly contaminated at some latitudes, it will show a peak in those latitudes while the other classes will show a dip, as a result of the contamination in the denominator of the number ratio. Plotting sky densities instead is lees instructive since the most noticeable trend is a drop in density toward the Galactic plane, for all classes, that can be explained by the combination of Galactic extinction and the steep function of galaxy or AGN number counts with apparent magnitude. 

The distribution between the different AGN types is indeed largely independent Galactic latitudes but only for $|l|{\gtrsim}30^\circ$, showing a strong deviation below that. The bottom panel in Fig. \ref{fig:fractions_lat} shows the effect of removing point sources with high {\it Gaia} proper motions, specifically sources with both ps-score$>0.8$ and PM/PM$_{\rm error} >4$. This cleaning appears to partially alleviate the contamination by stars to the high-z AGN and blazar classes close to the Galactic plane. To limit this contamination further, we eliminated candidates that are both close to the Galactic plane (|Galactic latitude| $<30\deg$) and have MIR colours typical of stars ($W1-W2<0.3$). This cleaning produces slightly flatter class distribution in the 2 central bins%  shown in the bottom panel in Fig. \ref{fig:fractions_lat}. 
We note that this final selection is still contaminated by stars in regions with |Galactic latitude| $<30\deg$, especially in the high-z and Blazar classes. Outside this region the distribution of candidates among different AGN classes is largely independent of Galactic latitude, as expected. 

In section \ref{section:sample_comparison} we compare our variability-selected AGN candidates with other catalogues, considering only our candidates that fulfil the cleaning criteria described above.

\section{Comparison to other AGN samples}\label{section:sample_comparison}

Below we compare the populations of AGN candidates selected by variability (removing possible contaminants as described in Sec. \ref{section:cleaning}) to those selected by other methods, namely colour-based selections and X-ray detections. 

\subsection{Gaia-unWISE colour-based AGN selection (C75 catalog)}
Survey averaged photometry is often used to find AGN candidates through colour selections. For instance, \citet{shu2019} use a random forest algorithm fed on photometric and astrometric optical data from {\it Gaia} and photometric MIR data from WISE to produce an all-sky large catalogue of Quasar candidates. Here we compare our variability-based selection method to the colour selection method of \citet{shu2019} using their 'C75' catalogue, which has its probability threshold fitted to produce a completeness of 75\%.

In an area of overlap with our sample defined by $|gal_{lat}| \geq 20, -28\leq dec\leq 15.5$ there are 318,887 AGN candidates in our sample and 865,589 AGN candidates in the C75 catalogue. Cross-matching both samples by sky coordinates with a tolerance of 1\farcs5 returns 298,717 matches.  

\begin{figure}
    \centering
    \includegraphics[width=0.48\textwidth]{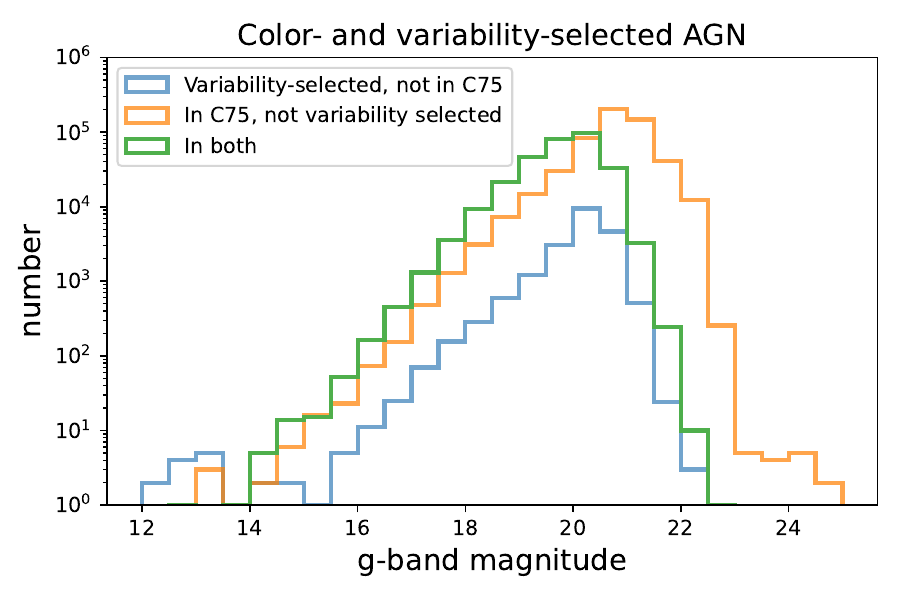  }
    
    \caption{{\it Gaia} g-band magnitude distribution of AGN candidates in our ZTF selection and in the {\it Gaia}-unWISE selection C75. The distribution of sources in both catalogues is plotted in green, of sources in the ZTF catalogue and not in the C75 catalogue in blue, and sources in the C75 catalogue and not in the ZTF catalogue in yellow. Evidently the largest source of incompleteness in the ZTF catalogue is its limited depth.}
    \label{fig:ZTF_GUA}
\end{figure}

Figure \ref{fig:ZTF_GUA} shows the $g$-band magnitude distribution of the matched candidates in green, the candidates that appear only in our variability-selected  catalogue in blue and the C75 candidates that were not selected by our method in yellow, where all samples are restricted to the overlap area defined above. To build this plot, we cross-matched the {\it Gaia}-unWISE C75 catalogue to Pan-STARRS DR-1 to obtain optical magnitudes in the same photometric system as our sample. Evidently, a significant fraction of the difference between our selection and the C75 catalogue is caused by the shallower magnitude limit of the ZTF data we used. For a more limited range of magnitudes, $g=14-20$, our method recovers roughly three quarters of the C75 candidates (164,122 of 221,367), dropping to 50\% at $g=20-20.5$ and only about  1\% at dimmer magnitudes. 

Making a cut in $14<g<20$ mag to control the effect of incompleteness due to the sensitivity of ZTF leaves 164,122 matches, 57,222 un-matched C75 sources and 5,433 un-matched ZTF sources. 

The ZTF footprint misses 13\% of the area coverage due to gaps between the chips \footnote{https://iopscience.iop.org/article/10.1088/1538-3873/ab4ca2}, while we have rejected additional objects falling close to these gaps or that otherwise raise SExtractor flags in the reference images or in the difference images. To establish how much our classifications differ we searched all un-matched C75 candidates, with magnitude $14<g<20$ mag, that appear in our master catalogue obtaining only 6,479 matches, of which only 3,220 had light curves of sufficient quality to produce a classification. Therefore, of the 57,222 unmatched C75 candidates, over 54,000 were not in our selection simply because we did not classify them. This unclassified fraction corresponds to 24\% of the C75 candidates with $14<g<20$ mag. Considering only sources for which we did attempt a classification, our method classified only 2 \% of the C75 candidates in this magnitude range as something other than an AGN. These remaining 3220 objects had the following types in our classifier: 2594 as non-variable galaxy, 222 as non-variable star, 258 as CV/Nova, 124 as SNe and the rest as other variable stars. 

On the other hand, our variability selection includes 5,433 objects with $14<g<20$ that are not considered in C75. Cross matching this sample to the DESI  DR1 \citep{DESI_DR1} spectroscopic catalogue \footnote{zall-pix-iron.fits} returns 1,348 objects with good quality spectra (selected by the DESI label $\tt ZWARN=0$). The DESI pipeline classifications for these objects are QSO (957/1,348 = 71\%), galaxy (295/1,348 = 22\%) and the remainder 7\% are classified as stars. The redshift distribution of the QSOs is bimodal, with one peak centred at $z=0.15$ and the other at $z=2.2$, with similar numbers in both peaks. The redshift distribution of the galaxies is concentrated toward low values, with over 80\% of the objects having $z<0.3$. It is possible that some of these local galaxies contain lower luminosity AGN that could be recovered with detailed spectral fitting, which will be the focus of future work.

\subsection{X-ray AGN selection with eROSITA in eFEDS}
The eROSITA Final Equatorial Depth Survey \citep[eFEDS][]{Brunner2022} is a deep, continuous area survey in soft X-rays produced with eROSITA, where \citet{Salvato2022} has identified the optical counterparts to X-ray sources and provided a classification into Galactic and extra-galacticc classes, and \citet{Liu2022} has obtained the X-ray properties of the AGN candidates. In this section we compare our classifications in the eFEDS area. 

There are 22,079 AGN in the eFEDS AGN catalogue described in \citet{Liu2022}. We note that for the present work we used catalogue v17.6. Cross-matching these sources, using the coordinates from the optical counterparts, with PanSTARRS to obtain psf-photometry $g$-band magnitudes comparable to those used throughout this paper, we find 17,604 matches. Of these,  12,772 have $g>20$, too faint for reliable single-epoch ZTF imaging, so for these sources our selection is very incomplete. Noting also that our selection loses about 24\% of the area due to gaps and other SExtractor flags, we end up with  classifications for only 3,824 sources of the eFEDS AGN catalogue. 

Most, i.e. 2,995/3,824 or 79\%, of the eFEDS AGN that were included in our analysis were classified as AGN of different types by the variability classifier. The majority of the remaining eFEDS AGN were classified in our analysis as non-variable galaxies (766/3824 or 20\%), with the few remaining objects classified as stars and stellar transients of various types. 

\begin{figure*}
    \centering
    \includegraphics[width=0.45\textwidth]{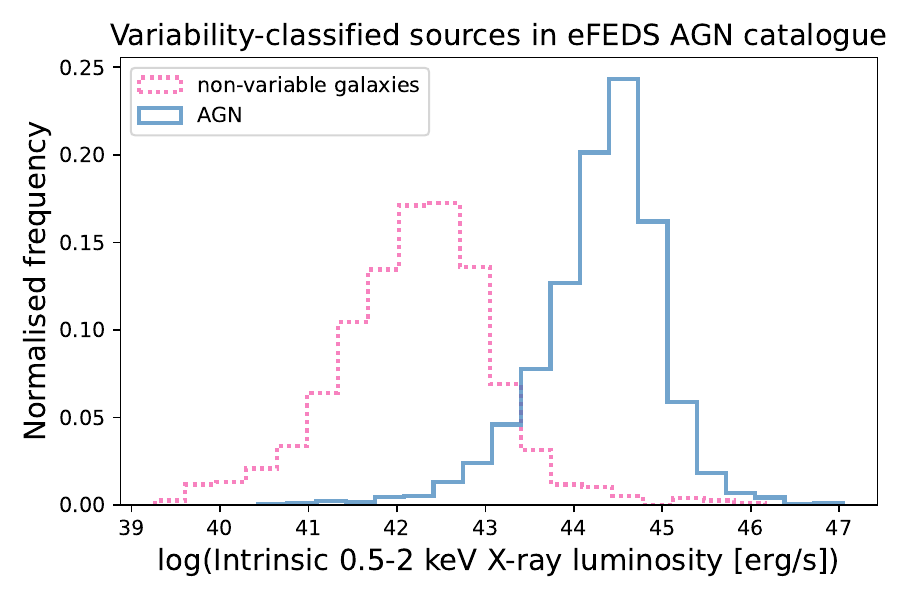}
     \includegraphics[width=0.45\textwidth]{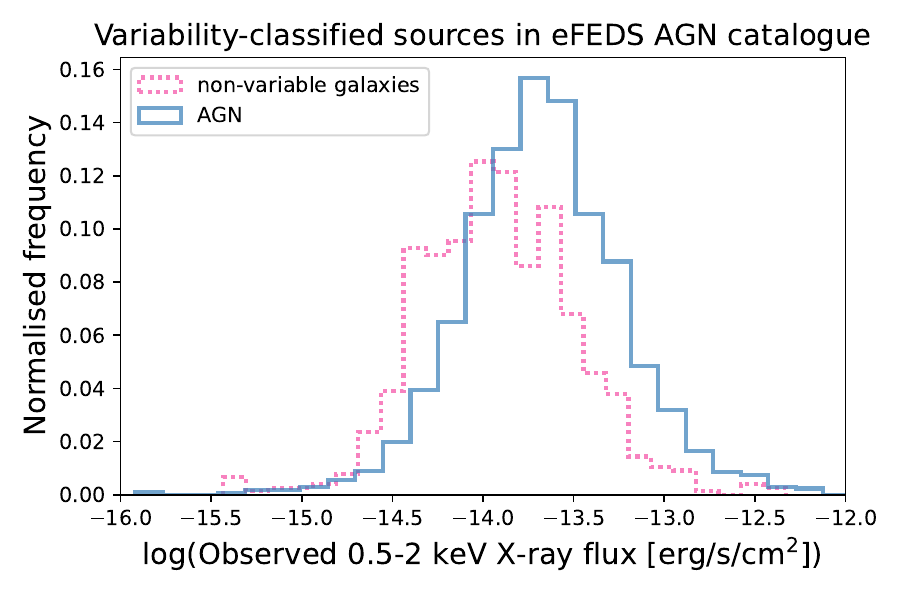}

   \caption{Comparison of X-ray properties of the eFEDS AGN included in our analysis. In all panels the eFEDS AGN that are also classified as AGN by our variability-based selection are shown in blue and those where our classification resulted in a non-variable galaxy type are shown in pink. The histograms are normalised to the total number of objects in each set, which is four times larger for the AGN class. All the X-ray properties are collected from the literature and are explained in \citet{Liu2022}.}
    \label{fig:efeds}
\end{figure*}

In Fig. \ref{fig:efeds}, we show the comparison of the X-ray properties of the eFEDS AGN from \citet{Liu2022} that are identified by our method as AGN in blue and identified by our method as non-variable galaxy, in pink. All the X-ray properties are taken from \citet{Liu2022} as no modelling of X-ray data was performed in the present work. 
The most striking difference between the samples is in the X-ray luminosity, as shown in the first panel, where almost all the higher X-ray luminosity objects were identified as AGN by our classifier, i.e. 90\% of our AGN candidates have log(L$_{0.5-2 \rm{keV}})>43.4$ and the median luminosity is log(L$_{0.5-2 \rm{keV}})=44.4$. On the other hand, eFEDS AGN identified as non-variable galaxies by our classifier had lower X-ray luminosities, with 90\% of them having log(L$_{0.5-2 \rm{keV}})<43.2$, with a median of log(L$_{0.5-2 \rm{keV}})=42.2$. This difference of 2 dex in median luminosity is unlikely to be a result of the differences in flux, as the X-ray flux distributions are very similar for eFEDS AGN classified as AGN or as non-variable galaxies by our classifier, as shown in the second panel. 

We note that among the eFEDS AGN that we classify as non-variable galaxies, 281/766=38\% have an X-ray luminosity in the 0.5--2 keV band above $3\times10^{42}$ so they should correspond to AGN. We note also that of the eFEDS AGN with $L_X>3\times10^{42}$ that had a variability-based classification, 2966/3247=91\% were classified as AGN. In Fig. \ref{fig:efeds_DRWtau} we show normalized histograms of the variability feature DRW $\tau$ for objects in the eFEDS area. The distributions show that the efeds AGN that are classified as non-variable galaxy have a slight excess of sources with DRW $\tau \sim 100-1000$, compared to the rest of the non-variable galaxies in the area, that do not have an X-ray counterpart. These might correspond to genuinely variable AGN that the classifier selects as non-variable galaxy possible due to a greater weight of other features, such as colour and extension.  As the orange histogram in Fig. \ref{fig:efeds_DRWtau} has over 61.000 non-variable, non-X-ray detected galaxies, and the pink histogram only has 766 X-ray detected, non-variable galaxies, this recovery is not simply a matter of reclassifying the more variable objects as AGN as this would increase the contamination. Instead, a new classifier could be built, in the future, including a low-luminosity AGN label.

For our variability-selected AGN candidates inside the eFEDS area (i.e., within the region with 90\% of its nominal exposure), there are 3,872 objects. Of these, 1,127 do not match any object in the eFEDS AGN catalogue, using the optical coordinates of the later for cross-matching. The vast majority of these unmatched variability-selected AGN do not have an X-ray counterpart within 10$\arcsec$ in the eFEDS point source catalogue either, so they are not missing a match due to an incorrect association of optical counterparts to the X-ray sources. Figure \ref{fig:efeds_area} shows the sky location of the eFEDS AGN in blue and our 1,127 unmatched AGN candidates in orange.  
We compared the distribution of features of the unmatched variability-selected AGN to the ones with eFEDS counterparts, finding similar distributions for number of epochs, light-curve length and class probability given by the classifier. The distribution of variability amplitudes quantified through the DRW$\sigma$ parameter are very similar for both samples, where the unmatched candidates have only a slightly higher proportion of low-variability sources. The distributions of the characteristics timescales quantified through the DRW$\tau$ parameter, are again very similar, with the unmatched sources having a slightly larger proportion of short timescales, consistent with the sampling rate. In either case, the matched and unmatched sources in general show very similar variability properties. The distributions of these properties are shown in Appendix \ref{app:Pvar}.   The most significant difference is observed between the $g-$band magnitudes of the matched and unmatched AGN candidates, shown in the right panel of Fig.~\ref{fig:efeds_area}, with the unmatched sample (orange) being $\sim$0.5 mag dimmer on average than the matched sample (green). 

\begin{figure*}
    \centering
    \includegraphics[width=0.49\textwidth]{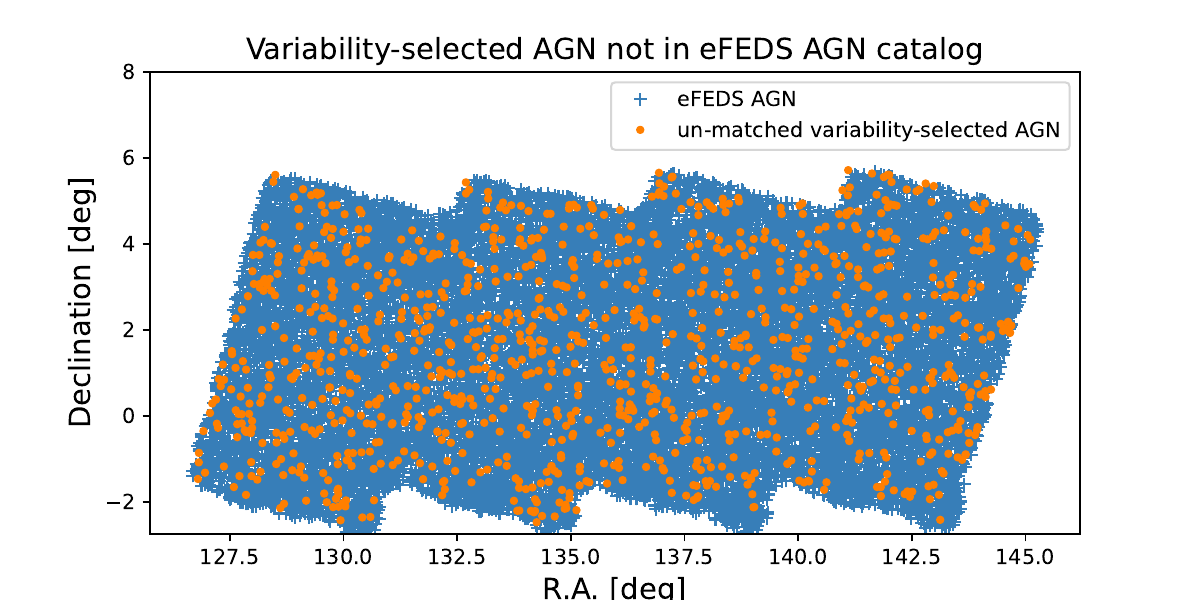}
    \includegraphics[width=0.49\textwidth]{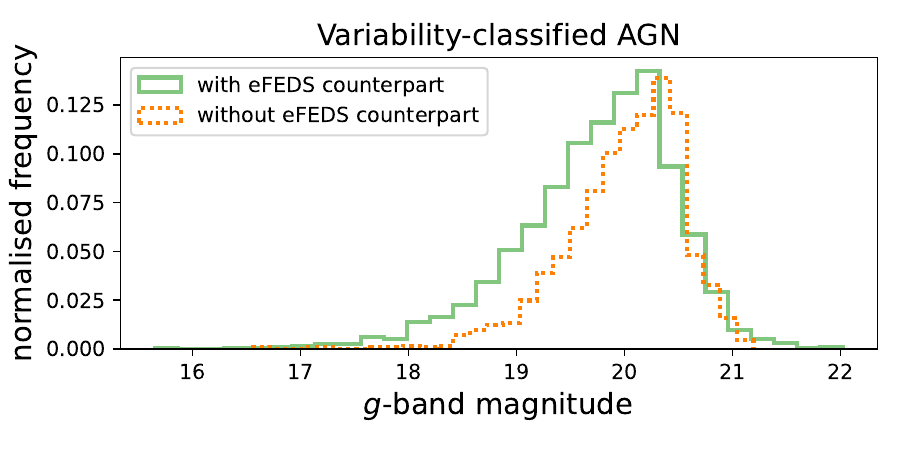}

   \caption{Left panel: Sky location of the eFEDS AGN within the area with over 90 \% coverage, in blue. Variability-selected AGN candidates that do not match the eFEDS AGNs and are within this area are marked in orange.  Right panel: the distribution of $g-$band magnitudes for both sets, where variability-classified AGN in the eFEDS field with a match in the eFEDS AGN catalogue is plotted in green and without a match in orange. }
    \label{fig:efeds_area}
\end{figure*}

Since the eFEDS area is well covered by the SDSS DR16 Quasar catalogue, we used this to investigate the nature of our matched and unmatched sources. Of the 1,127 variability-selected AGN candidates that lack eFEDS AGN counterparts, 432 appear in the SDSS DR16 catalogue of Quasar properties of \citet{Wu2022}. On the other hand, of the 2,995 variability-selected AGN that do have an X-ray counterpart, 1,203 appear in this catalogue. We note that in both samples, the fraction that appear in the SDSS Quasar catalogue is similar (38\% vs 40\%). We also find similar distributions of black hole masses and Eddington ratios for variability candidates with and without eFEDS counterparts, but higher median redshifts for the variability-selected AGN candidates without eFEDS counterparts, i.e. $z_{med}=1.41$ for sources with eFEDS counterparts vs $z_{med}=1.95$ for sources without. The difference between the matched and un-matched, variability-selected AGN will be discussed in detail in a follow-up paper. 

\section{Low redshift AGN}
One of our goals is to select low luminosity AGN at low redshift, where the host galaxies appear extended, complicating the extraction of accurate lightcurves. In Figs. \ref{fig:std}, \ref{fig:tau} and \ref{fig:Pvar} we show that the light curves presented here are better behaved than the lightcurves obtained for the same objects with psf-photometry on the un-subtracted images. In this section we show that the low-z AGN candidates indeed include extended objects and not simply the point source AGN in the local Universe. The ps-score feature was calculated by \citet{Tachibana18} to quantify the extension of astronomical sources. Our low-z AGN candidate list covers a range of ps-scores, with two concentrations: 42\% of candidates have ps-score $<0.2$ (i.e. extended) and 30\% have ps-score $>0.8$ (i.e. point like). The mid-z AGN candidates in contrast have mainly point like structures where only 0.5\% have ps-score$<0.2$  and 95\% have ps-score $>0.8$. 

\begin{figure}
    \centering
    \includegraphics[width=0.49\textwidth]{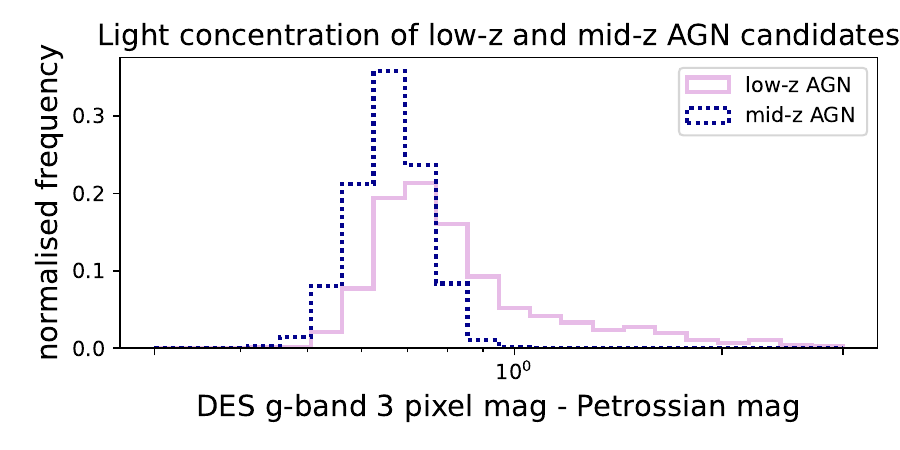}

   \caption{Distribution of differences between small 3-pixel apertures magnitudes and Petrossian magnitudes, obtained from DES DR2 for the low-z and mid-z AGN candidate samples.   }
    \label{fig:extension}
\end{figure}

We also consider the light concentration by comparing the magnitude from a small aperture to the total (i.e. Petrossian) magnitude.  Figure \ref{fig:extension} shows this magnitude difference ($\Delta M$) for all the mid-z and low-z AGN candidates that appear in the Dark Energy Survey DR2 \citep{Abbot2021}, from where we obtained these values. The blue, dashed histogram, containing the mid-z AGN candidates shows this $\Delta M$ for mostly point sources. The distribution of concentrations of the low-z AGN candidates, in pink, is partly consistent with the point sources but it contains a long tail towards higher values, or lower concentrations. Up to 21\% of the low-z AGN sample that appears in the DES DR2 catalogue has $\Delta M > 1$, while almost no mid-z AGN candidate reaches this difference.   In Appendix \ref{app:lcs} we show a few examples of lightcurves of low-z AGN candidates that correspond to extended and/or host dominated galaxies. There we compare our lightcurves to their DR counterparts and show the images of the selected galaxies. 
The low-z AGN candidates that appear in low stellar mass galaxies are largely confirmed as type I AGN via their optical spectrum. This analysis was carried out by \citet{Bernal2024}, who also compares the success rate of AGN selection in low mass galaxies to other works using optical variability to select AGN candidates. In brief, the AGN candidates selected here that coincided with low stellar mass galaxies ($M_*<2\times 10^{10}M_\odot$) had significant broad Balmer lines in 182 out of 188 cases with archival spectra, and 74\% of the candidates within the eROSITA-DE sky had X-ray counterparts in their DR1 \citep{Merloni2024}. Other works following similar procedures have found much lower rates of both, broad Balmer lines and X-ray counterparts among their variability-selected low mass AGN candidates. A detailed comparison can be found in Sec. 6.2 of \citet{Bernal2024}.  

\section{Summary and conclusions}\label{section:sumary}

 The aperture photometry on ZTF reference-subtracted (i.e. difference) images produces more accurate light curves and the calibration proposed here produces more accurate errors than the ZTF light curves distributed in the data releases. The difference is particularly important for low-redshift galaxies, both quiescent and active, which show similar levels of variability in the DR light curves but very different levels of variability in our DI-Ap  light curves. 
 
 This caveat must be considered when using ZTF data release light curves to search for AGN in low redshift ($z<0.5$) galaxies through variability, which will often be spurious. For example, in non-active nearby galaxies, 98\% of the objects satisfy the standard variability test {\tt Pvar}>95\% when studied with the DR11-psf light curves while, for the same objects, only 7\% satisfy this criterion when studied with our DI-Ap  light curves. For comparison, using the same {\tt Pvar} criterion, low redshift AGN show a variable fraction of 99\% with the DR light curves, similar to the non-active galaxies, while the DI-Ap  light curves produce a variable fraction of 76\%, much greater than that found with this method for the non-active galaxies. It is therefore also important to use these DI-Ap  light curves when attempting to quantify the variability of low redshift AGN, where the DR11-psf light curves are contaminated by non-intrinsic variations and their errors are underestimated. 
 
 As shown in Sec. \ref{section:features_comparison}, apart from the amplitude of fluctuations, other variability features like the characteristic timescale of variations quantified by the DRW-$\tau$ parameter also depend on the type of light curve used. The DI-Ap  light curves separate much more cleanly cases where true characteristic timescales are recovered from those where the only noticeable timescale is the sampling rate, and produce far fewer intermediate characteristic timescales in low-redshift and mid-redshift AGN (see Fig. \ref{fig:tau}), as well as for high-redshift AGN.  

We further used our DI-Ap  light curves of all objects in our $>$8,000 deg$^{2}$ region of study to gauge their usefulness for classification purposes. We applied the same methodology, classes and training sets as detailed in \citetalias{Sanchez-Saez2023} but replaced the DR11-psf  light curves that were used in that work by our DI-Ap  light curves. Our classifier follows a hierarchical approach (BHRF), where we first separate sources as variable or non-variable and then as transient, stochastic, or periodic objects, and in a third level, we further resolve each of these classes. We have four AGN classes, including three bins of redshift and Blazars. The macro-averaged scores of the BHRF model are 0.58, 0.76, 0.60, for the precision, recall, and F1-score, respectively. However, when considering only the AGN classes and grouping them as a single class, we obtain a precision, recall, and f1-score of 1.00, 0.96, and 0.98, respectively. This implies that we expect a purity close to 100\% and a completeness close to 96\% in our AGN selection. We attempted to construct light curves for the 42,020,693 sources in the ZTF-ID list, which are in an area limited approximately by $-29<dec<15$ and $|Gal_{lat}|>20$. This method produced classifications for 39,772,280 light curves. After removing duplicated sources, using a radius of 1\farcs5, we ended up with a catalogue of 341,938 AGNs, including the four different AGN classes. Figure \ref{fig:numclass} summarizes the number of candidates obtained for each class, without removing duplicated sources. 

When compared to the classifier based on DR11-psf light curves of \citetalias{Sanchez-Saez2023} we note an improved performance detecting AGN in low redshift galaxies, some of which were classified as non-variable galaxies with the DR11-psf light curves. This improvement is in line with what was expected since the main improvement of the new light curves is the removal of the non-variable host galaxies contaminating the variable nuclear component, which dilutes the variability in the DR light curves. The correct identification of AGN candidates in low redshift galaxies is important for the detection of supermassive black holes in the low-mass range. In \citet{Bernal2024} we show that this classification, when cross matched with a sample of low-stellar mass, low-redshift galaxies, consistently produced bona fide low-mass AGN, detectable through their broad emission lines in the optical spectrum, and with a high fraction of detections in the X-ray band.  

The variability-based selection of AGN candidates produced in this work is similar to the sample obtained through multi-wavelength colour selection of \citet{shu2019} (C75 catalogue ) in that, of the objects we did classify, the majority appear in their catalogue and in that few objects from C75 where classified as something other than an AGN by our classifier. These 3220 miss-matched classifications represent only 2\% of the C75 objects that where also classified by variability, and that are in the $14<g<20$ magnitude range. They are mostly identified as non-variable galaxies by our variability-based classification. On the other hand, 5,433 variability-selected AGN candidates in this magnitude range were not included in the C75 catalogue. Given their colours we expect them to correspond to low-luminosity AGN where the host galaxy dominates the colour. Their nature will be confirmed with spectroscopic data, which at least in part will be obtained in the near future through the ChANGES \citep{Bauer2023} survey of 4MOST.   

We compared our variability-selected AGN candidates to the X-ray selected AGN candidates based on eROSITA data in the eFEDS field. 
The eFEDS AGN that were included in our analysis were almost exclusively classified as extragalactic sources, where 80\% corresponded to AGN classes and 20\% to non-variable galaxies. These two populations are well separated by their X-ray luminosity, although they have similar X-ray fluxes. We conclude from here that eFEDS AGN identified as non-variable galaxies in our variability classifier correspond to a different population of weak AGN, where the optical variability is negligible and the X-ray luminosity is on average 2 orders of magnitude lower than the eFEDS AGN classified as AGN through their variability as well.

Conversely, about a quarter of the variability-selected AGN in the eFEDS fields did not have an X-ray counterpart, although about 40\% of each sample (with and without X-ray matches) appear in the SDSS DR16 Quasars catalogue. These un-matched candidates are on average 0.5 mag dimmer in the optical than the variability-selected AGN with eFEDS counterparts. This might explain their lack of X-ray detections, if the optical to X-ray flux ratios are sufficiently varied.
Completing the spectroscopic coverage of AGN candidates in this region, both selected by X-ray detections and through optical variability will therefore help to constrain the diversity of X-ray to optical ratios in AGN.  

\begin{acknowledgements}

{ We thank the anonymous referee for their helpful comments and suggestions, which helped to improve this paper.} We acknowledge support from Millennium Science Initiative Program NCN$2023\_002$ (PA, PL, SR), Millennium Science Initiative, AIM23-0001 (PA, FEB), FONDECYT Regular 1241422 (PA, PL, FEB, BS), 1241005 (FEB, PA), ANID-Chile BASAL project CATA FB210003 (FEB), and CAV, CIDI N. 21 U. de Valparaíso, Chile (PA).

Based on observations obtained with the Samuel Oschin Telescope 48-inch and the 60-inch Telescope at the Palomar Observatory as part of the Zwicky Transient Facility project. ZTF is supported by the National Science Foundation under Grants No. AST-1440341 and AST-2034437 and a collaboration including current partners Caltech, IPAC, the Weizmann Institute for Science, the Oskar Klein Center at Stockholm University, the University of Maryland, Deutsches Elektronen-Synchrotron and Humboldt University, the TANGO Consortium of Taiwan, the University of Wisconsin at Milwaukee, Trinity College Dublin, Lawrence Livermore National Laboratories, IN2P3, University of Warwick, Ruhr University Bochum, Northwestern University and former partners the University of Washington, Los Alamos National Laboratories, and Lawrence Berkeley National Laboratories. Operations are conducted by COO, IPAC, and UW.

This work has made use of data from the European Space Agency (ESA) mission
{\it Gaia} (\url{https://www.cosmos.esa.int/gaia}), processed by the {\it Gaia}
Data Processing and Analysis Consortium (DPAC,
\url{https://www.cosmos.esa.int/web/gaia/dpac/consortium}). Funding for the DPAC
has been provided by national institutions, in particular the institutions
participating in the {\it Gaia} Multilateral Agreement.

This research used data obtained with the Dark Energy Spectroscopic Instrument (DESI). DESI construction and operations is managed by the Lawrence Berkeley National Laboratory. This material is based upon work supported by the U.S. Department of Energy, Office of Science, Office of High-Energy Physics, under Contract No. DE–AC02–05CH11231, and by the National Energy Research Scientific Computing Center, a DOE Office of Science User Facility under the same contract. Additional support for DESI was provided by the U.S. National Science Foundation (NSF), Division of Astronomical Sciences under Contract No. AST-0950945 to the NSF’s National Optical-Infrared Astronomy Research Laboratory; the Science and Technology Facilities Council of the United Kingdom; the Gordon and Betty Moore Foundation; the Heising-Simons Foundation; the French Alternative Energies and Atomic Energy Commission (CEA); the National Council of Humanities, Science and Technology of Mexico (CONAHCYT); the Ministry of Science and Innovation of Spain (MICINN), and by the DESI Member Institutions: www.desi.lbl.gov/collaborating-institutions. The DESI collaboration is honored to be permitted to conduct scientific research on I’oligam Du’ag (Kitt Peak), a mountain with particular significance to the Tohono O’odham Nation. Any opinions, findings, and conclusions or recommendations expressed in this material are those of the author(s) and do not necessarily reflect the views of the U.S. National Science Foundation, the U.S. Department of Energy, or any of the listed funding agencies.
\end{acknowledgements}

\bibliographystyle{aa}
\bibliography{bibliography.bib}

\begin{appendix}
\section{Distributions of other properties}
\label{app:Pvar}
\begin{figure}
    \centering
    \includegraphics[width=0.4\textwidth]{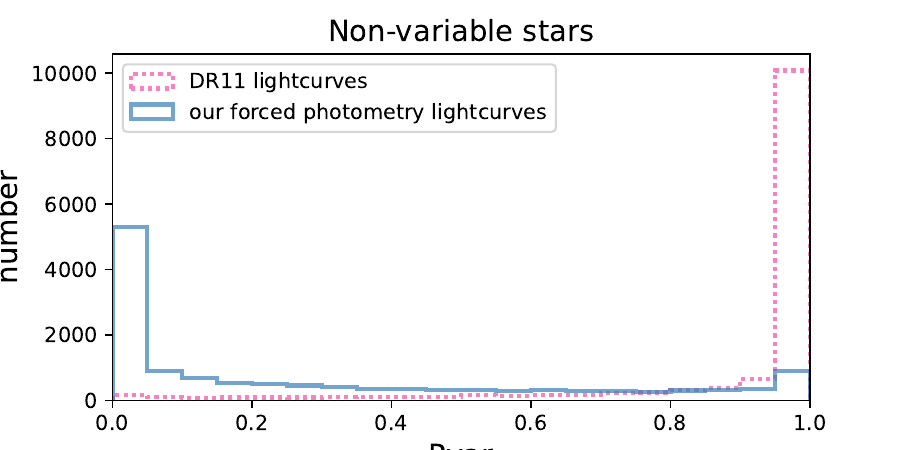}
     \includegraphics[width=0.4\textwidth]{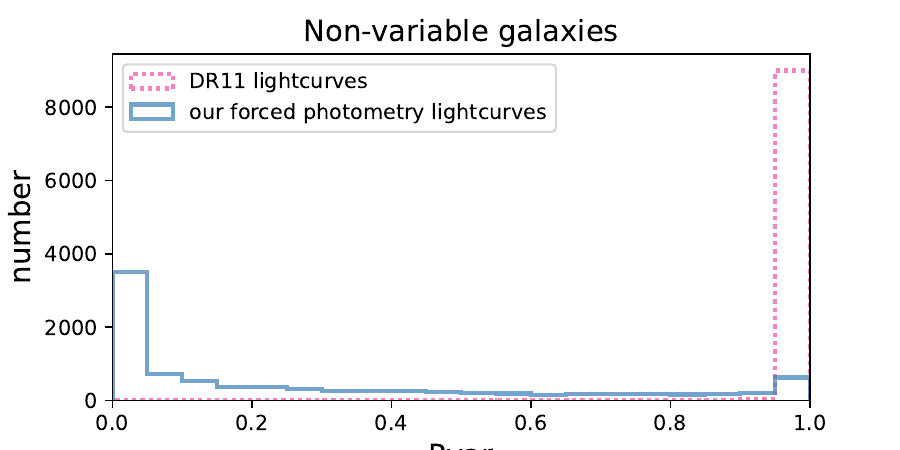}
    \includegraphics[width=0.4\textwidth]{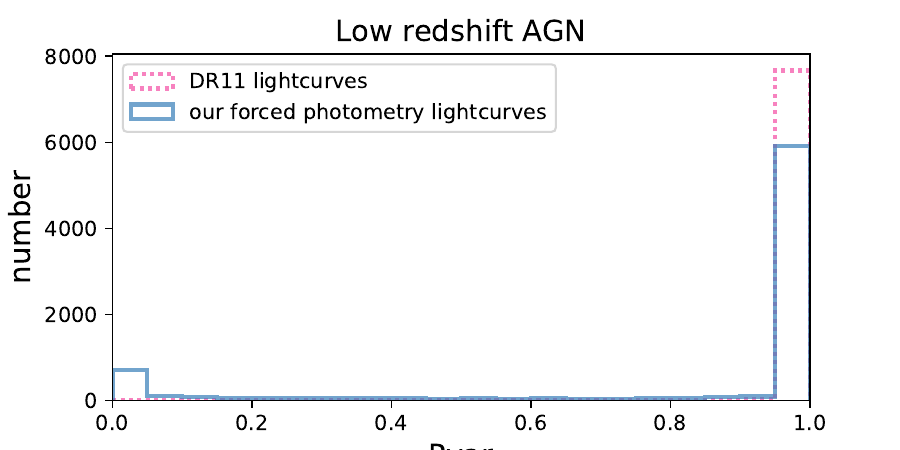}
     \includegraphics[width=0.4\textwidth]{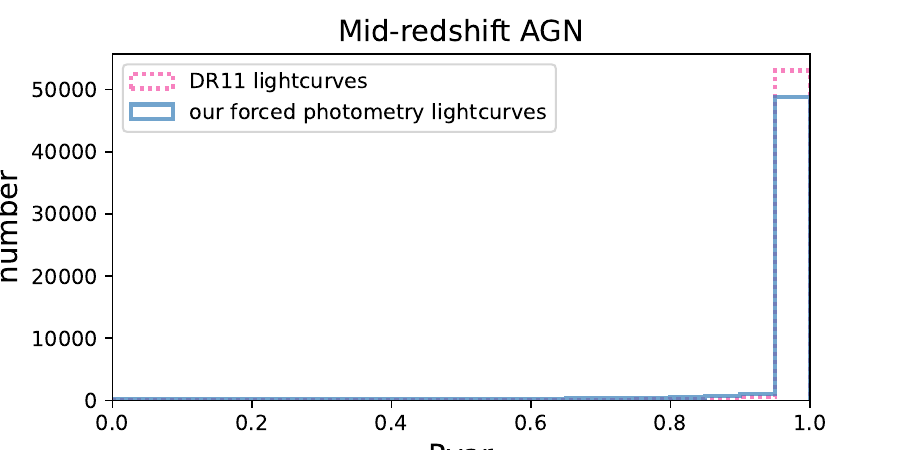}

   \caption{Distribution of the {\tt Pvar} values of the light curves and associated errors obtained from our custom-made photometry (blue, solid line) and directly from the ZTF data release DR11-psf light curves (pink, dashed line) for four types of objects in the labelled set: non variable stars, non-variable galaxies, and low redshift ($z<0.5$) AGN and mid-redshift AGN ($0.5<z<3$).}
    \label{fig:Pvar}
\end{figure}

Figure \ref{fig:Pvar} shows the distribution of the probability that an object is variable measured by \texttt{Pvar}, when using the DR11-psf  light curves and when using our DI-Ap  light curves for the same objects. The panels show the difference in  \texttt{Pvar} distributions for non-variable stars and galaxies, which should cluster around \texttt{Pvar=0} and for variable objects (low-z and mid-z AGN) which should cluster around \texttt{Pvar=1}.

Figure \ref{fig:efeds_area} shows the variability properties of variability-selected AGN candidates with and without counterparts in the eFEDS catalogue of X-ray-detected AGN.

\begin{figure}
    \centering
    \includegraphics[width=0.49\textwidth]{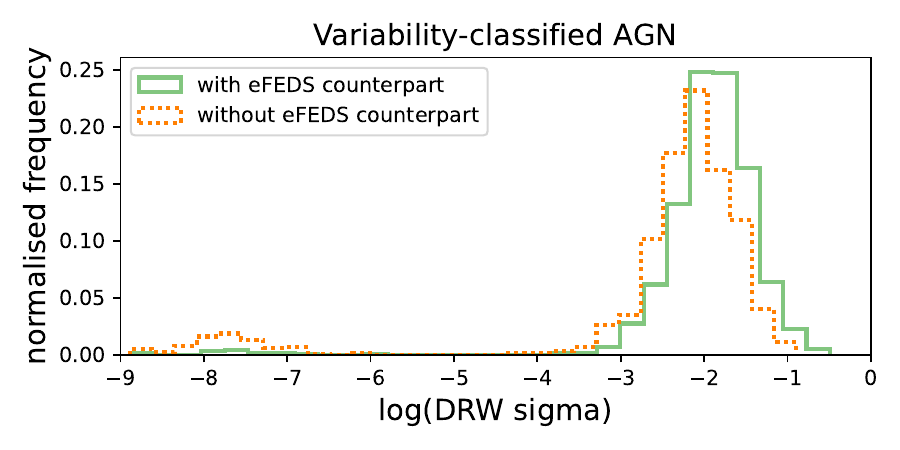}
    \includegraphics[width=0.49\textwidth]{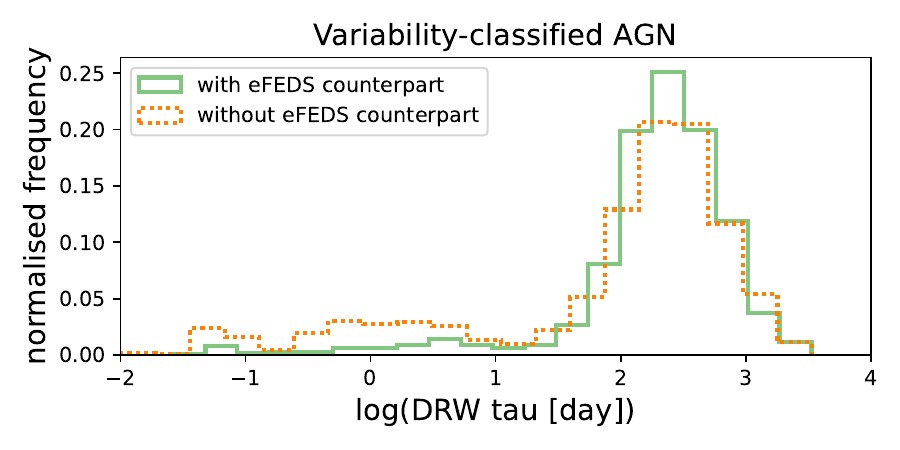}

   \caption{Variability-classified AGN in the eFEDS field, with a match in the eFEDS AGN catalogue (green) and without a match (orange). Top panel: distribution of variability amplitudes quantified through the DRW$\sigma$. Bottom panel: distribution of the characteristics timescales quantified through the DRW$\tau$ parameter. }
    \label{fig:efeds_area}
\end{figure}

\begin{figure}
    \centering
    \includegraphics[width=0.45\textwidth]{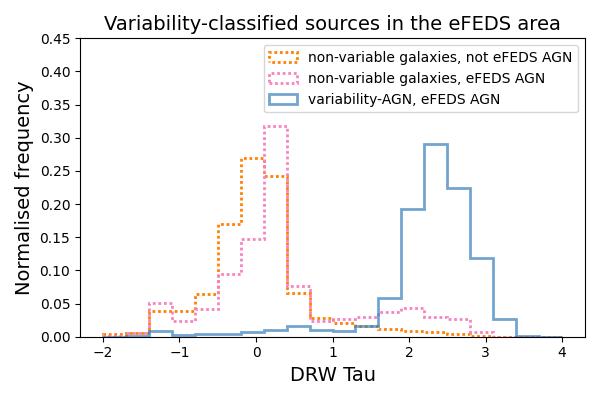}

   \caption{Comparison of the variability feature DRW $\tau$ eFEDS AGN included in our analysis, classified as AGN (2,995 objects, blue), and classified as non-variable galaxy (766 objects, pink), togetehr with the non-variable galaxies that do not have an X-ray counterpart although they are in the eFEDS area (61,640 objects). The histograms are normalized by area to facilitate the comparison.}
    \label{fig:efeds_DRWtau}
\end{figure}

Figure \ref{fig:efeds_DRWtau} shows the variability properties DRW $\tau$ of eFEDS AGN classified as AGN, as non-variable galaxies, and other non-variable galaxies that do not have and X-ray counterpart even though they are in the eFEDS field. 

\section{Comparison of lightcurves from extended, low-z AGN}
\label{app:lcs}
\begin{figure}
    \centering
    \includegraphics[width=0.24\textwidth]{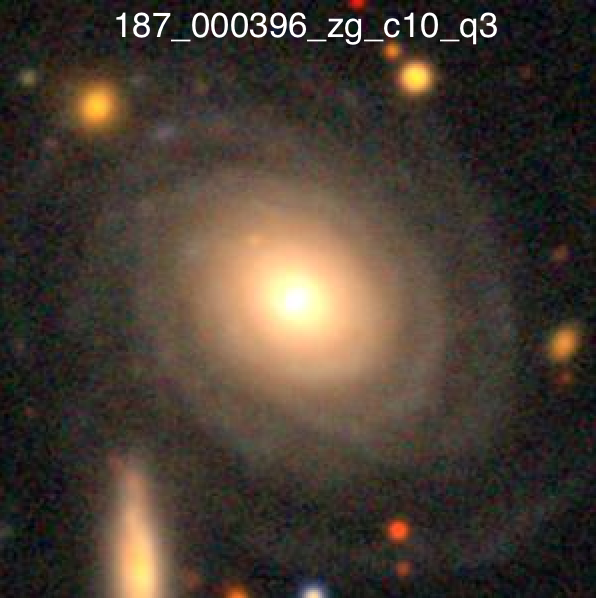}
    \includegraphics[width=0.24\textwidth]{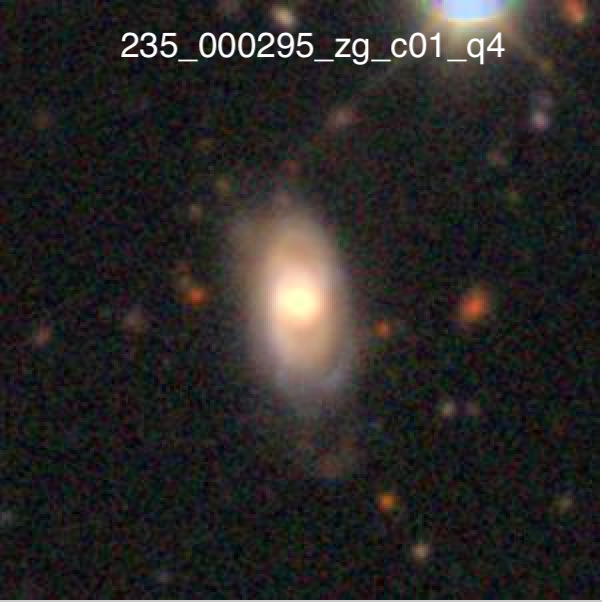}
    \includegraphics[width=0.24\textwidth]{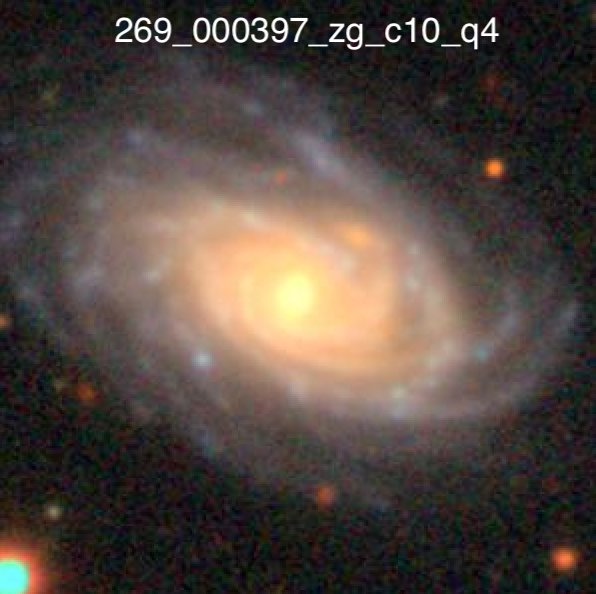}
    \includegraphics[width=0.24\textwidth]{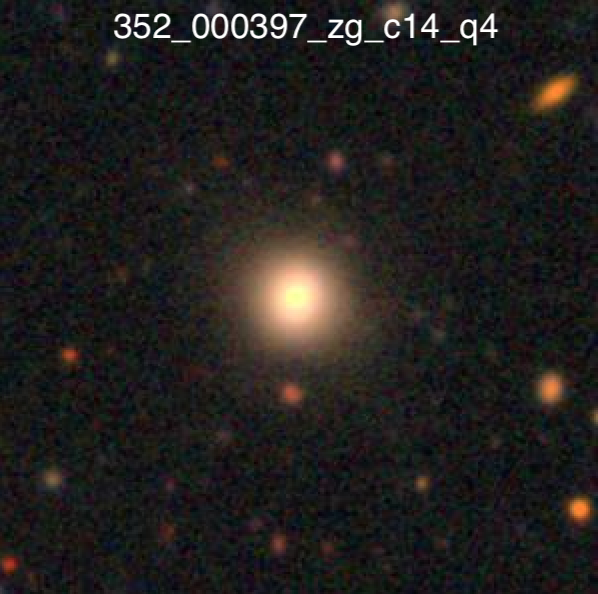}

   \caption{Images of the 4 galaxies included in Fig. \ref{fig:lcs_DR_FP} obtained from the DES DR2 colour images database. The stamps have a size of 1 arcminute per side. }
    \label{fig:lcs_DR_FP_images}
\end{figure}

\begin{figure}
    \centering
    \includegraphics[width=0.49\textwidth]{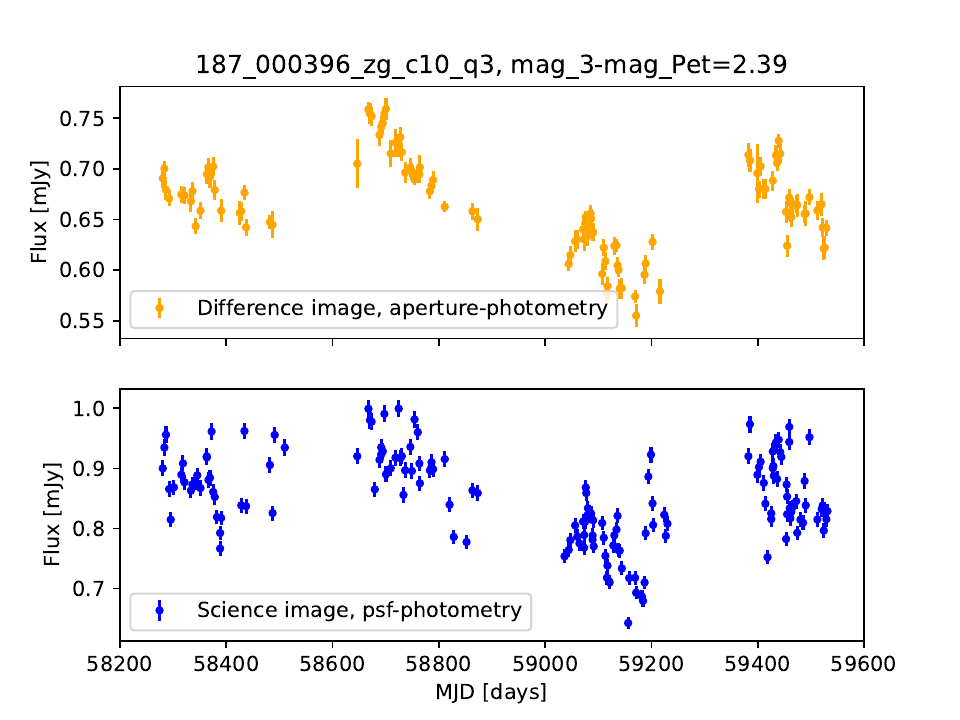}
\vspace{-0.5cm}    
\includegraphics[width=0.49\textwidth]{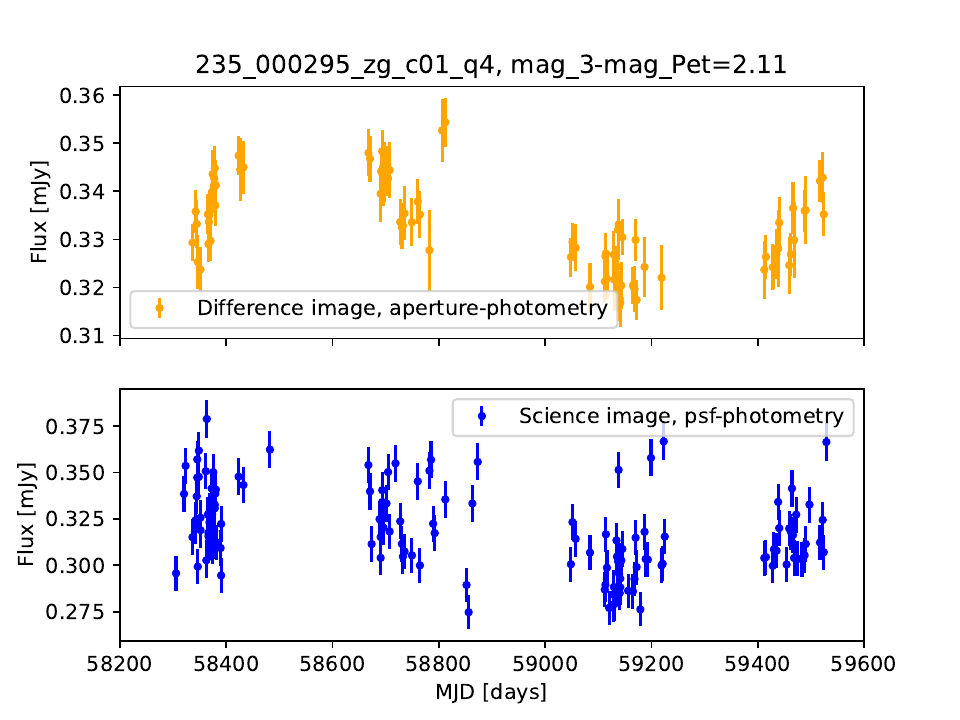}
\vspace{-0.5cm}
\includegraphics[width=0.49\textwidth]{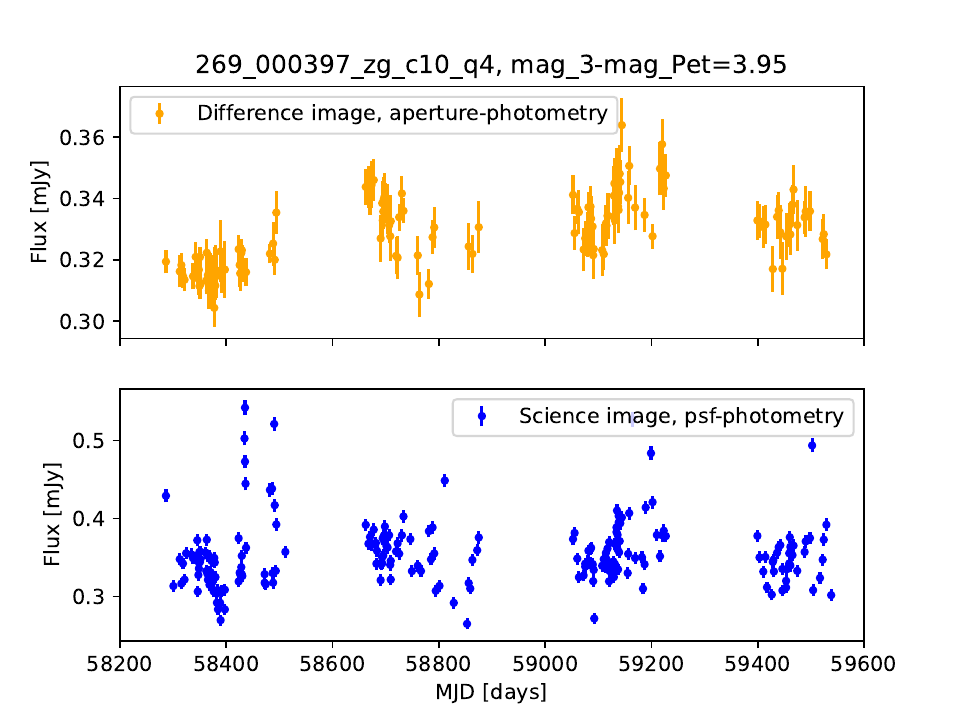}
\vspace{-0.5cm}
\includegraphics[width=0.49\textwidth]{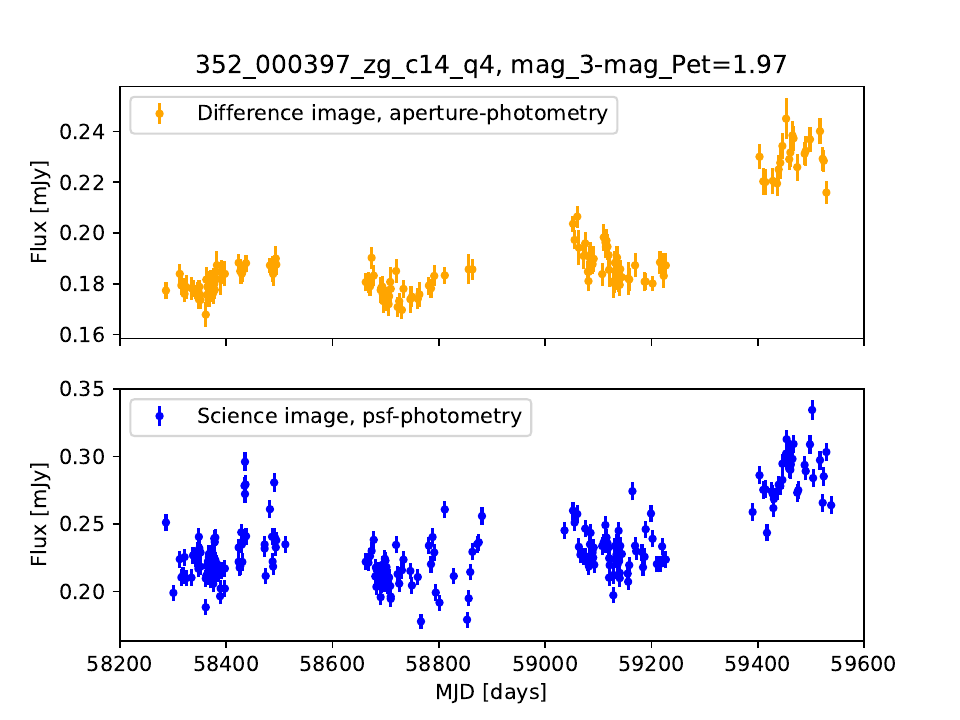}
\vspace{-0.1cm}
   \caption{Comparison between our DI-Ap  lightcurves in orange and the DR psf-photometry lightcurves in blue, for 4 extended galaxies classified as low-z AGN.   }
    \label{fig:lcs_DR_FP}
\end{figure}

In Fig. \ref{fig:lcs_DR_FP} we show some example lightcurves of low-z AGN candidates, obtained with the DI-Ap photometry presented here, in orange, and from the ZTF DR psf-photometry directly on the science images, in blue. These objects were chosen among the low-z AGN that had DES photometry in different apertures available and showed large differences between the 3-pixel aperture magnitude and the Petrossian magnitude. The values of these differences are noted in the title of each image along with the object identifier from our photometry. They were also chosen to represent groups with different extensions and concentrations. The DR lightcurves in the plot were cleaned retaining only epochs with \textit{catflags}=0 and \textit{limitmag}>20 to remove bad nights and epochs with image processing issues. The corresponding images are shown in Fig. \ref{fig:lcs_DR_FP_images}.

The DR lightcurves exhibit issues that complicate the distinction between low-redshift AGN and non-active galaxies. Specifically, some flux points show spurious large deviations that are not accounted for by their small error bars. These same outliers appear in the lightcurves of inactive galaxies, causing metrics like Pvar and Excess Variance to misclassify them as variable. In the case of low-z AGN, this over-prediction of short-timescale variability—driven by these outliers—leads to incorrect values for characteristic timescales, as seen in the Damped Random Walk $\tau$ statistic.

\end{appendix}
\end{document}